\documentclass[
nofootinbib,
reprint,
amsmath,
amssymb,
aps,
superscriptaddress,
preprintnumbers,
prd
]{revtex4-2}
\usepackage{hyperref}
\hypersetup{colorlinks=true, citecolor=blue, urlcolor=blue, linkcolor=blue}
\usepackage[T1]{fontenc}
\DeclareRobustCommand{\VAN}[3]{#2}
\let\VANthebibliography\thebibliography
\def\thebibliography{\DeclareRobustCommand{\VAN}[3]{##3}\VANthebibliography}

\usepackage{comment}
\renewcommand{\comment}[1]{}

\def\thebibliography{\DeclareRobustCommand{\VAN}[3]{##3}\VANthebibliography}
\usepackage{graphicx}	
\usepackage{amsmath}	

\usepackage{amssymb}	
\usepackage{bm}		

\usepackage{ae,aecompl}
\usepackage{newtxtext,newtxmath}
\usepackage{xurl}

\usepackage{dcolumn}    		
\usepackage{aas_macros}
\usepackage{tipa} 
\usepackage[table]{xcolor}

\newcommand{\eftc}{\vec{c}}
\newcommand{\kmaxfitcs}{k_{\eftc}}
\newcommand{\kmax}{k_{\rm UV}}
\newcommand{\thetamin}{\theta_{\rm UV}}

\newcommand{\refeq}[1]{Eq.~(\ref{eq:#1})}

\newcommand{\reffig}[1]{Fig.~\ref{fig:#1}}          
          
\newcommand{\reftab}[1]{Tab.~\ref{tab:#1}}          
\newcommand{\refsec}[1]{Sec.~\ref{sec:#1}}          
\newcommand{\refapp}[1]{App.~\ref{app:#1}}
          
\renewcommand{\selectlanguage}[1]{} 

\date{\today}

\begin{document}

\title[Towards the Two-Loop EFTofLSS in Galaxy Lensing Surveys]{Towards the Two-Loop EFTofLSS in Galaxy Lensing Surveys}

\author{Evan Saraivanov}
\email[First author: ]{evan.saraivanov@stonybrook.edu}
\affiliation{Department of Physics and Astronomy, Stony Brook University, Stony Brook, NY 11794, USA }
\affiliation{C.N. Yang Institute for Theoretical Physics, Stony Brook University, Stony Brook, NY 11794, USA }

\author{Henrique Rubira}
\email[Second author: ]{henrique.rubira@lmu.de}
\affiliation{University Observatory, Faculty of Physics, Ludwig-Maximilians-Universit\"at, Scheinerstr. 1, D-81679 München, Germany}
\affiliation{Kavli Institute for Cosmology Cambridge, Madingley Road, Cambridge CB3 0HA, UK}
\affiliation{Centre for Theoretical Cosmology, Department of Applied Mathematics and Theoretical Physics
	University of Cambridge, Wilberforce Road, Cambridge, CB3 0WA, UK}

\author{Vivian Miranda}
\affiliation{C.N. Yang Institute for Theoretical Physics, Stony Brook University, Stony Brook, NY 11794, USA }

\author{Tim Eifler}
\affiliation{Department of Astronomy and Steward Observatory, University of Arizona, 933 N Cherry Ave, Tucson, AZ 85719, USA}
\affiliation{Department of Physics, University of Arizona,  1118 E Fourth Str, Tucson, AZ, 85721-0065, USA}

\begin{abstract}
{
Extracting cosmological information from Stage IV weak lensing surveys requires non-linear modelling of the matter power spectrum that is accurate across a broad range of scales and redshifts and robust to baryonic feedback. We forecast the application of the two-loop effective field theory of large-scale structure (EFTofLSS) to Roman Space Telescope, carefully considering parameterization, scale cuts, and priors. We develop neural network emulators for the two-loop integrals, allowing rapid evaluation of the likelihood. 
Weak lensing demands a continuous-in-redshift description of the EFT, potentially introducing tens of nuisance parameters. We address this by calibrating the counterterm redshift evolution against the \textsc{Euclid Emulator 2} and accounting for the residual freedom in redshift with spline functions. A principal component analysis of the free parameters reduces the dimensionality to a few degrees of freedom that the data can constrain. Next, we calibrate the priors on those degrees of freedom by using a suite of hydrodynamical simulations. 
We forecast the $S_8$ constraints as a function of scale cuts, showing that the two-loop EFT with Roman cosmic shear provides unbiased $S_8=\sigma_8\sqrt{\Omega_{\rm m}/0.3}$ constraints with relative errors of about $0.9\%$ and $1.4\%$ when allowing for $5\%$ and $1\%$  contamination from ultraviolet modes, respectively. The two-loop EFT improves the scale reach beyond the one-loop EFT and non-linear dark matter-only models when baryonic effects are included. This framework provides a robust path for extracting small-scale information from future cosmic shear data.
}
\end{abstract}

\maketitle

\section{Introduction}

Stage IV galaxy surveys such as the Nancy Grace Roman Space Telescope (Roman)~\cite{dore2019wfirstessentialcosmologyspace,akeson2019widefieldinfraredsurvey}, the Legacy Survey of Space and Time (LSST)~\cite{Ivezi__2019}, the Dark Energy Spectroscopic Instrument (DESI)~\cite{https://doi.org/10.5281/zenodo.7858207}, and the Euclid mission~\cite{laureijs2011eucliddefinitionstudyreport} will deliver unprecedented high-resolution and high-precision measurements of the late-time Universe. LSST and Euclid have recently commenced survey operations, DESI has already produced multiple data releases~\cite{desicollaboration2026datarelease1dark,Abdul_Karim_2025}, and Roman is scheduled to launch in the near future. Fully exploiting the constraining power of these datasets requires theoretical predictions for non-linear clustering of comparable precision across a wide range of scales and redshifts.

High-resolution $N$-body simulations provide a well-controlled description of non-linear gravitational clustering on small scales, contingent upon the physical assumptions encoded in the simulation. However, they are computationally expensive and therefore limited in the range of cosmological parameter space that can be directly explored. For $\Lambda$CDM and minor extensions beyond it, such as $w$CDM, large simulation suites are available, such as \textsc{Quijote}~\cite{Quijote_sims}, and power spectrum emulators such as \textsc{Euclid Emulator 2} (\textsc{EE2})~\cite{2021} and \textsc{Mira Titan}~\cite{Heitmann_2016} interpolate across parameter space.
Beyond fully simulation-based approaches, semi-analytic prescriptions based on the halo model~\cite{Seljak_2000,Ma_2000,Peacock_2000} have been widely adopted for their computational efficiency and empirical accuracy deep into the non-linear regime~\cite{Takahashi_2012,Mead_2021}. 
An additional complication arises from baryonic physics. Hydrodynamical processes such as gas cooling, star formation, and feedback from active galactic nuclei (AGN) modify the matter distribution on small and intermediate scales. Capturing these effects requires hydrodynamical simulations, which are substantially more expensive than dark matter-only simulations and introduce further modeling uncertainties. In practice, baryonic effects are often incorporated through phenomenological parameterizations calibrated on specific simulation suites, adding nuisance parameters that must be marginalized over~\cite{Schneider:2015wta,Mead:2020vgs,Mead_2021,Aric__2021,Arico:2023ocu}. 

An alternative strategy is to parametrize the effects of small-scale physics affects large-scale observables analytically within the framework of the effective field theory of large-scale structure (EFTofLSS)~\cite{Baumann_2012,Carrasco_2012,Carrasco_2014,Konstandin_2019}.
The EFTofLSS extends the standard perturbation theory used to solve the continuity and Euler's equations by introducing additional parameters, namely Wilson coefficients or counterterms. These counterterms parameterize the backreaction of the unresolved short-scale physics in the ultraviolet (UV) modes onto the infrared (IR) modes. This is achieved without referencing an explicit model of baryonic feedback.
In a practical analysis, these parameters must be marginalized over, which generally degrades the constraining power on the fundamental cosmological parameters.

The EFTofLSS has been predominantly applied to spectroscopic galaxy surveys~\cite{d_Amico_2020,D_Amico_2021,Zhang_2022,Simon_2023,chudaykin2025priorsscalecutseftbased,ivanov2025fullshapeanalysissimulationbasedpriors,ramirezsolano2024modelingparametercompressionmethods}, where galaxy clustering is analyzed in redshift space. In this context, the signal is localized within relatively narrow redshift bins, allowing the EFT counterterms to be treated as independent parameters at each effective redshift.
For weak lensing, the situation is qualitatively different. Gravitational lensing is sensitive to the integrated matter distribution along the entire line of sight between the observer and the source galaxies. As a result, the observables receive contributions from a broad range of redshifts, and the EFT coefficients must be modeled as evolving functions of redshift rather than as single effective numbers. Several works have explored possible prescriptions for this evolution, for example, by parameterizing deviations from analytically motivated scaling relations~\cite{Foreman:2015uva}.

More recently, the EFTofLSS has also been applied to weak lensing analyses. \citet{damico2025cosmologicalanalysisdes3times2pt} employed the one-loop EFT matter power spectrum for the lensing component of their $3\times2$pt analysis using DES year three data, while modeling galaxy clustering using the one-loop EFT of biased tracers~\cite{Desjacques_2018}. They reported a modest degradation in constraining power, together with a small shift in the inferred matter clustering parameter $S_8 = \sigma_8\sqrt{\Omega_{\rm m}/0.3}$. 
In addition, \citet{derose2025lensingcounternarrativeeffective} implemented the one-loop EFT of galaxy shapes~\cite{Vlah:2019byq,Chen:2023yyb} and demonstrated that, with a careful treatment of ultraviolet contributions, it is possible to obtain improved constraints on $S_8$.

In this manuscript, we develop a framework for applying for the first time the two-loop EFTofLSS matter power spectrum to weak lensing. We construct a neural-network emulator to efficiently evaluate the two-loop matter power spectrum with sub-percent accuracy. The EFT counterterms and their redshift dependence are calibrated by first performing a fit to dark matter-only simulations, capturing their leading behavior, and then introducing cubic B-spline functions with free amplitudes to account for residual corrections from non–dark matter physics on small scales.
To control the dimensionality of the parameter space, we perform a principal component analysis (PCA) and keep only the dominant modes that most significantly impact weak lensing observables, reducing the number of additional nuisance parameters. The priors on the PCA amplitudes are calibrated using publicly available hydrodynamical simulations that include baryonic feedback, ensuring that the counterterms can effectively absorb baryonic contributions to the total matter power spectrum.

We demonstrate this framework using a Roman real-space cosmic shear likelihood, fully specifying the PCA basis to allow for a complete forecast analysis. Fisher forecasts are used to obtain constraints on $S_8$, considering fiducial datavectors constructed from noiseless dark matter-only models as well as noiseless models that include baryonic feedback. We compare the constraining power and parameter bias obtained using the linear matter power spectrum, a non-linear dark matter-only power spectrum computed using \textsc{EE2}, and both the one-loop and two-loop EFT predictions. This comparison is performed as a function of the wavenumber $\kmax$, which defines the corresponding angular scale cuts based on the maximal allowed UV contamination.
We find that the two-loop EFT allows the inclusion of smaller angular scales without significant loss of constraining power. 
In contrast, the one-loop EFT performs comparably to dark matter-only non-linear prescriptions and does not offer the same improvement in scale reach, particularly when baryonic effects are present.

The structure of the paper is as follows. In \refsec{EFT}, we review the two-loop EFT framework, highlighting our treatment of the redshift evolution of the counterterms. \refsec{dataanalysis} describes the Roman survey specifications adopted in this work, introduce the cosmic shear observables, and present the methodology used to translate wavenumber scale cuts into angular cuts, along with the statistical framework employed in the analysis. We present our results in \refsec{results} and summarize our conclusions in \refsec{conclusion}.

\section{The Two-loop EFT of Large-Scale Structure} \label{sec:EFT}

We review the two-loop effective field theory of large-scale structure for matter in \refsec{reviewEFT}. In \refsec{eft_evolution}, we explain how we treat the EFT counterterms and their redshift dependence. 

\subsection{Review of EFTofLSS at two loops} \label{sec:reviewEFT}

The effective field theory of large-scale structure~\cite{Baumann_2012,Carrasco_2012,Carrasco_2014,Konstandin_2019} provides an analytic description of matter and biased tracers in the mildly non-linear regime. This is achieved by introducing an effective stress tensor for the matter fluid that accounts for imperfect fluid corrections, such as velocity dispersion and viscosity, as well as effective pressure perturbations induced by small-scale physics. The terms in the effective stress tensor contain all operators allowed by the equivalence principle, suppressed by powers of the non-linear scale $k_{\rm NL}$.

The EFT matter power spectrum up to two loops can be writen as
\begin{equation}\label{eq:eftspectrum}
    P_{\rm EFT} = P_{\rm L} + (P_{1,{\rm SPT}} - P_{1,{\rm ct.}}) + (P_{2,{\rm SPT}} - P_{2,{\rm ct.}})\,,
\end{equation}
where $P_{\rm L}$ is the linear matter power spectrum, $P_{\ell,{\rm SPT}}$ is the $\ell$-th loop standard perturbative correction (see \cite{Bernardeau:2001qr} for a review), and $P_{\ell,{\rm ct.}}$ is the $\ell$-th loop counterterm. We use the Einstein--de-Sitter approximation (see  \cite{D_Amico_2021,Garny_2021,Garny_2022}) and rescale the loop contributions by powers of the linear growth factor $D(z)$ to account for their redshift dependence.

\begin{figure}[t]
    \centering
    \includegraphics[width=\columnwidth]{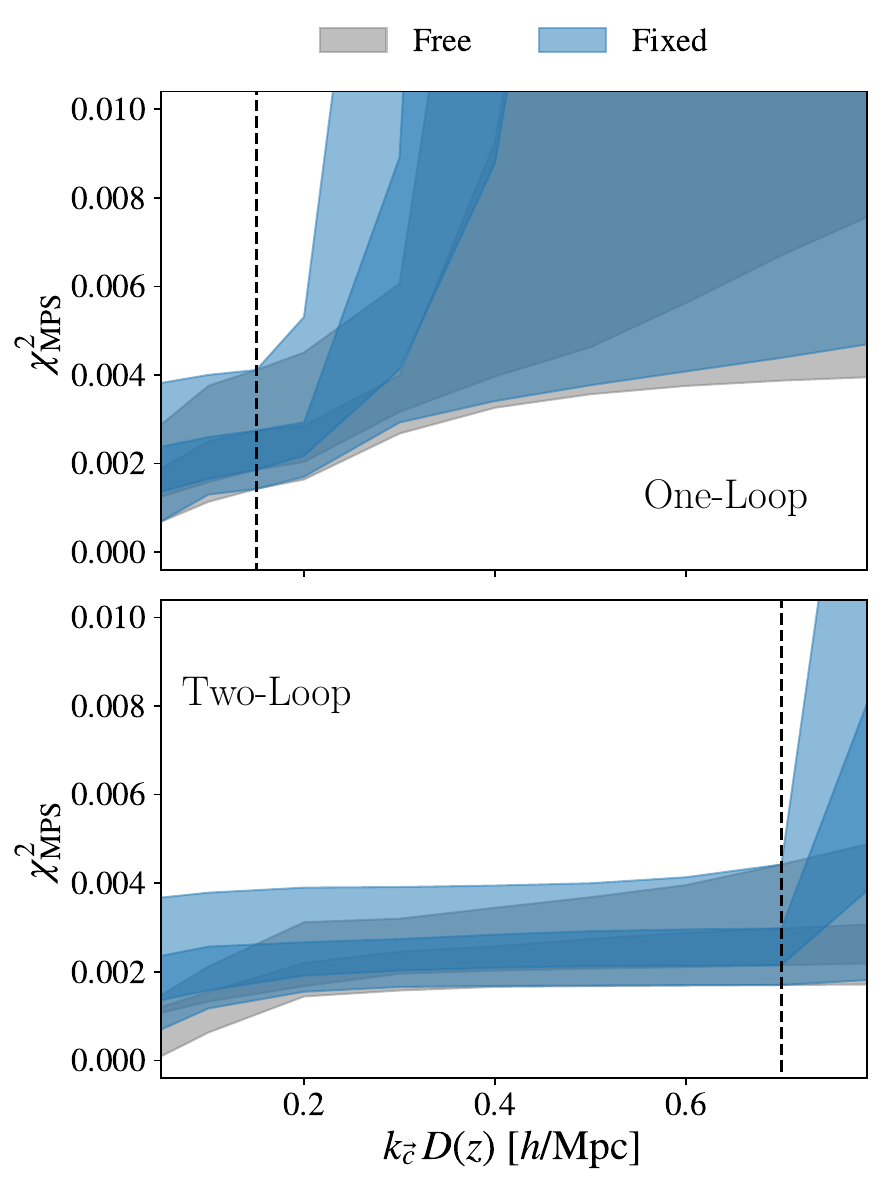}
    \caption{The $67\%$ and $95\%$ bands for $\chi^2_{\rm MPS}$ as a function of $\kmaxfitcs$, the largest $k$ used to fit the counterterms at $z=0$, rescaled by the linear growth factor, computed from  $899$ cosmologies and across $512$ redshifts. The one-loop is displayed in the upper panel and two-loop in the lower panel. The gray band shows the $\chi^2_{\rm MPS}$, defined in \refeq{chi2MPS}, if we allow the coefficients to take any value that minimizes the $\chi^2_{\rm MPS}$, while the blue bands show the result taking $\kmaxfitcs = 0.15/D(z)\,h/$Mpc for the one-loop and $\kmaxfitcs = 0.7/D(z)\,h/$Mpc for the two-loop (vertical dashed lines), and use it at another (lower) $\kmaxfitcs$. Using the chosen $\kmaxfitcs$, even if those are beyond the perturbative regime, gives a good fit to the power spectrum, as seen by the relatively stable $\chi^2_{\rm MPS}$.
    }
    \label{fig:coeffs_ee2ref}
\end{figure}

\begin{figure}[t]
    \centering
    \includegraphics[width=\columnwidth]{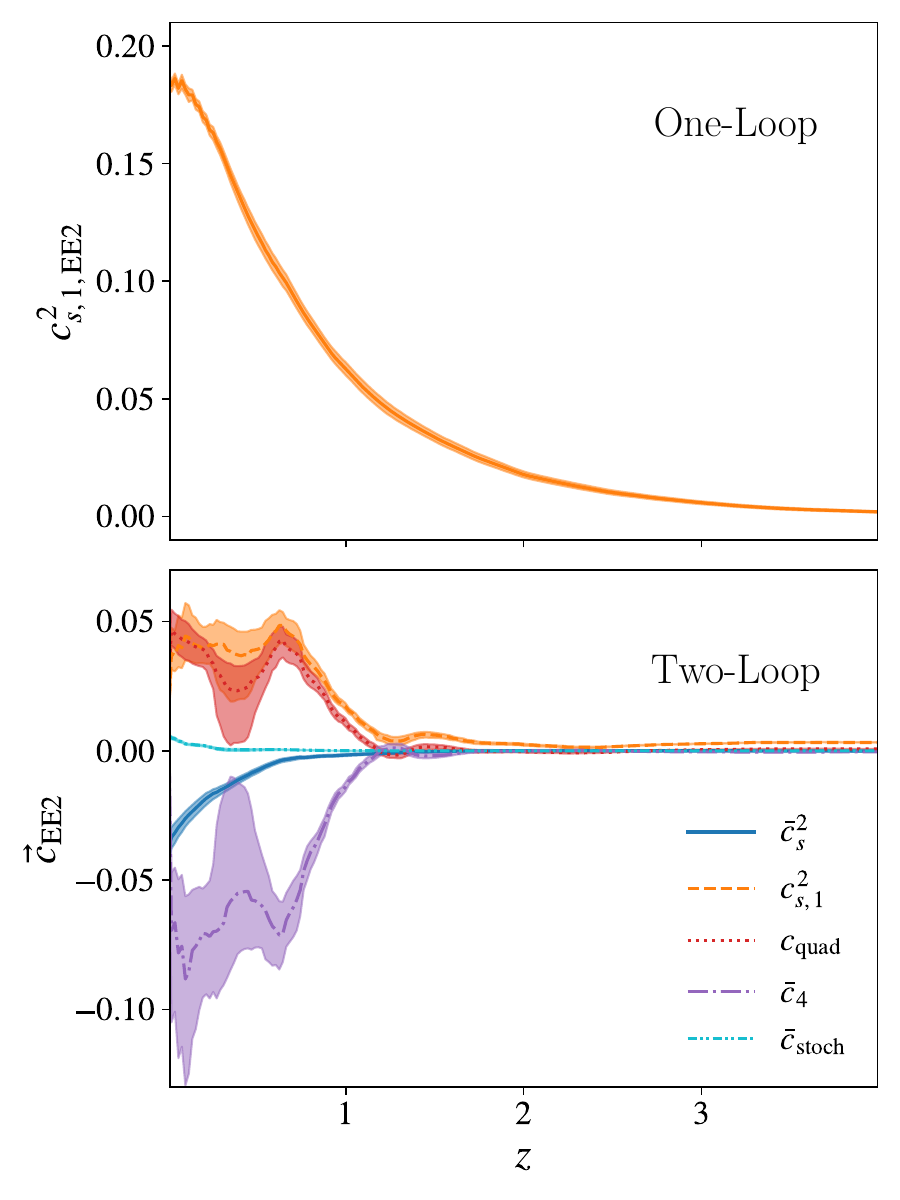}
    \caption{The best fit values of the counterterms fit to the \textsc{EE2} power spectrum as a function of redshift. The bottom panels display the values for $\eftc$ at two loops, while the top panel shows the single effective sound speed at one-loop. The solid line is the median across the $899$ cosmologies we used to train the emulators, and the shaded regions show the 67\% confidence interval across the $899$ cosmologies. These were fit at the values of $\kmaxfitcs$, as described in \reffig{coeffs_ee2ref}.}
    \label{fig:c_of_z}
\end{figure}

At one loop, the single counterterm is given by 
\begin{equation}\label{eq:ct1loop}
    P_{1,{\rm ct.}} = 2\pi^2 c_{s,1}^2 \frac{k^2}{k_{\rm NL}^2} P_{\rm L}\,,
\end{equation}
and at two loops, the counterterms are given by
\begin{equation}
\begin{split} \label{eq:ct2loop}
    P_{2,{\rm ct.}}  &= 
    2\pi^2 c_{s,2}^2 (z) \frac{k^2}{k_{\rm NL}^2} P_{\rm L}  + 2\pi^2 c_{s,1}^2 (z)\frac{k^2}{k_{\rm NL}^2} P_{1,{\rm SPT}} \\
    &\quad + \pi^4 \left([c_{s,1}^2(z)]^2 - 2c_4(z)\right) \frac{k^4}{k_{\rm NL}^4} P_{\rm L} \\
    &\quad + 2\pi^2 c_{\rm quad}(z)\frac{k^2}{k_{\rm NL}^2} P_{\rm quad} 
    + \pi^4 c_{\rm stoch}(z)\frac{k^4}{k_{\rm NL}^4}\,,
\end{split}
\end{equation}
with 
$P_{\rm quad} = \int d^3 q\, F_2(\mathbf{k}-\mathbf{q},\mathbf{q}) P_{\rm L}(|\mathbf{k}-\mathbf{q}|) P_{\rm L}(q)/(2\pi)^3$ as the counterterm for $\delta^2$, and $F_2$ being the usual second-order kernel \cite{Bernardeau:2001qr}. The coefficient $c_{s,1}^2$ is the effective sound speed of the matter fluid at one loop, while $c_{s,2}^2$, $c_4$, $c_{\rm quad}$, and $c_{\rm stoch}$ are the Wilson coefficients of the two-loop counterterms. Recently, \cite{Anastasiou:2025jsy} pointed out the existence of a few additional matter counterterms at two loops, which are suppressed compared to the five coefficients considered here. As discussed in \cite{Lewandowski:2014rca}, at one-loop order \refeq{ct1loop} is able to absorb baryonic effects into the matter power spectrum, while at two loops one would require an additional term~\cite{Braganca:2020nhv} proportional to the relative baryon–matter velocity. However, this contribution is suppressed relative to the other parameters. 

We implement the IR resummation technique \cite{Senatore:2017pbn} as outlined in~\cite{Garny_2022} (see Sec.~3.2 of \cite{Bakx:2025jwa} for full expressions).
We also implement the IR-safe integrands~\cite{Carrasco_2014}. The integrals are evaluated between $0.0001\,h/{\rm Mpc}$ and $50\,h/{\rm Mpc}$, with the counterterms completely absorbing the cutoff dependence of the loop integrals \cite{Foreman:2015lca, Konstandin_2019, Bakx:2025jwa}.

Calculating the two-loop power spectrum for a general cosmology is numerically challenging, as it involves a five-dimensional integral. Recently, \cite{Bakx:2025jwa} performed the first cosmology sampling of two-loop power spectra using a precomputed spectral basis introduced in \cite{Bakx:2024zgu}, demonstrating substantial gains over the one-loop result when considering three-dimensional galaxy maps. For alternative approaches to accelerating the two-loop computation, see \cite{Anastasiou:2025jsy}, which uses a massive-propagator basis, and \cite{Farakou:2025tuq}, which employs symbolic regression.
In this work, we emulate the two-loop power spectrum using a multi-layer perceptron with residual connections~\cite{he2015deepresiduallearningimage,zhong2024attentionbasedneuralnetworkemulators,saraivanov2024attentionbasedneuralnetworkemulators,zhu2025attentionbasedneuralnetworkemulators}. We find agreement at the sub-percent level for cosmological parameters allowed by Planck 2018 measurements~\cite{Planck:2018vyg} of the first acoustic peak of the CMB (TT, TE, and EE with $\ell<396$). Further details on the CMB prior used in this analysis can be found in \refsec{priors}, while details of the emulation procedure are presented in~\refapp{Loop_emulator}.

\subsection{Counterterms and their redshift evolution}\label{sec:eft_evolution}

\begin{figure}[t]
    \centering
    \includegraphics[width=\columnwidth]{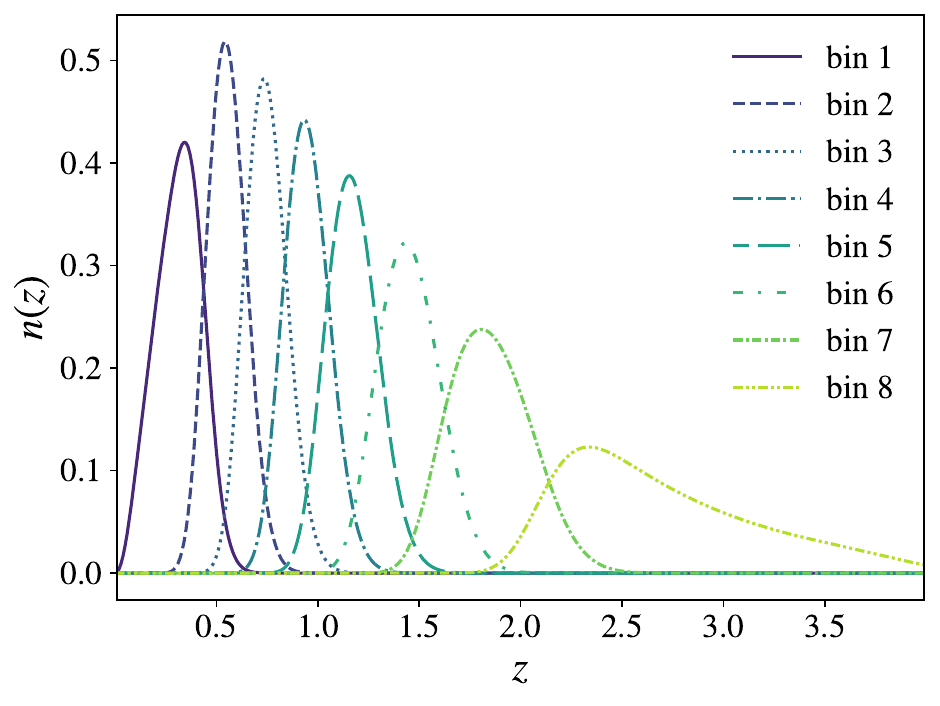}
    \caption{The redshift distribution of source galaxies used in our Roman cosmic shear forecast. The $n(z)$ in each of the $8$ redshift bins are normalized so that they integrate to unity.}
    \label{fig:roman_nz}
\end{figure}

The counterterms in the EFTofLSS framework parametrize unknown UV degrees of freedom, related both to non-linear dark matter dynamics and to baryonic physics, as well as canceling the cutoff dependence of the loop integrals. They are typically determined from data and then marginalized over in Bayesian analyses. Since these parameters are functions of redshift and depend on the cosmology, the standard procedure is to introduce a free set of parameters for each redshift bin (see e.g.~\cite{DESI:2024hhd}). Since we work with angular two-point functions in the context of photometric surveys, one must integrate over the line of sight, such that the full redshift dependence of these terms must be specified. One option, considered in \cite{damico2025cosmologicalanalysisdes3times2pt}, is to assume constant counterterms for each $\xi_{\pm}^{ij}$, which corresponds to the assumption that these terms do not vary significantly within each redshift bin. 

We pursue a different approach in this work. We calibrate the cosmology and redshift dependence using a suite of dark matter-only non-linear spectra $P_{\rm EE2}$ obtained with \textsc{Euclid Emulator 2} (EE2)~\cite{2021}, and use this as an ansatz around which we expand. This serves three purposes: first, to incorporate deviations from the fitted time dependence; second, to absorb residual errors from EE2; and third, to parametrize baryonic physics (see \refsec{priors} for our treatment of this and \refsec{baryons} for a forecast). While we opt to fully model the redshift dependence of the counterterms, we do not study the extent of its effect on lensing data. A rigorous comparison between the constant-in-redshift approach used in the one-loop analysis of \cite{damico2025cosmologicalanalysisdes3times2pt}, that considered one free counterterm parameter per $z$ bin, and our approach would provide insight into these modelling choices.

For the overall redshift dependence, we first perform a change of basis for the two-loop Wilson coefficients. We define $\bar{c}_s^2 = c_{s,1}^2+c_{s,2}^2$, $\bar{c}_4 = c_{s,1}^4-2c_4$ and $\bar{c}_{\rm stoch} = 10^{3} c_{\rm stoch}$, which capture the dominant degeneracies present in Eqs.~\eqref{eq:ct1loop} and \eqref{eq:ct2loop}. From this point on, we use the more compact notation $\eftc = \{\bar{c}_s^2, c_{s,1}^2, c_{\rm quad}, \bar{c}_4, \bar{c}_{\rm stoch}\}$ for the counterterms contributing to the two-loop power spectrum.
We then randomly sample $899$ cosmologies from a CMB first-peak posterior, within the bounds compatible with \textsc{EE2}, and evaluate the non-linear power spectrum at $513$ redshifts between $z=0$ and $z=10$. This dense sampling in redshift ensures that we obtain a near-continuous best-fit \textsc{EE2} prediction and allows us to safely use linear interpolation when the requested redshift sampling is coarser. We fit $\eftc$ to the \textsc{EE2} output up to the chosen $\kmaxfitcs(z)$, fixing the respective cosmology and obtaining the best-fit set $\eftc_{\rm \textsc{EE2}}$ by minimizing the function
\begin{equation}  \label{eq:chi2MPS}
\begin{split}
    \chi^2_{\rm MPS} &= \sum_{k_j \leq \kmaxfitcs } \left(\frac{P_{\rm EFT}(k_j,z_i) - P_{\rm EE2}(k_j,z_i)}{P_{\rm EE2}(k_j,z_i)}\right)^2 \\
    &\quad\quad\quad+ \lambda \left\rvert \eftc_{\rm \textsc{EE2}}(z_i) - \eftc_{\rm \textsc{EE2}}(z_{i-1}) \right\rvert^2 \,,
\end{split}
\end{equation}
where $\lambda = 0.01$ is a smoothing factor chosen to enforce the continuity of $\eftc_{\rm \textsc{EE2}}$ and to suppress rapid variations induced by any remaining degeneracies. For the point $z_{i=0}$, we perform a global minimization of the first term in $\chi^2_{\rm MPS}$ only.

Since $\kmaxfitcs$ also depends on redshift, we assume $\kmaxfitcs = \kmaxfitcs(z=0)/D(z)$ for its choice across different values of $z$. For a power-law model of the linear power spectrum, $P_{\rm L}(k) \propto k^{n}$, the non-linear scale $k_{\rm NL}(z)$ approximately scales as $1/[D(z)]^{2/(3+n)}$, such that this choice of $\kmaxfitcs$ is equivalent to setting $n = -1$.
Note that this $\chi^2_{\rm MPS}$ definition differs from the one used in \cite{Bakx:2025jwa}, which assigns different weights to different $k$ modes, as appropriate for three-dimensional spectroscopic maps, and leads to a growing $\chi^2_{\rm MPS}$ at low $k$ compared to \refeq{chi2MPS}. Our choice of $\chi^2_{\rm MPS}$, already considered in \cite{Konstandin_2019}, guarantees a high-quality fit across different values of $k$, as required for projected two-dimensional maps integrated over $k$. We emphasize that this $\chi^2_{\rm MPS}$ definition is used only for the calibration of the counterterms and not for the data analysis (see \refsec{dataanalysis}).

We show in \reffig{coeffs_ee2ref} the value of $\chi^2_{\rm MPS}$ as a function of $\kmaxfitcs$, rescaled by the growth factor. The lines indicate the $67\%$ and $95\%$ bands across the different redshifts and cosmologies used for the fit. The blue band shows $\chi^2_{\rm MPS}$ obtained by fitting all parameters at the indicated $\kmaxfitcs$, while the gray band corresponds to the value obtained when using the counterterms fitted at $0.15/D(z)\,h/{\rm Mpc}$ for the one-loop case and $0.7/D(z)\,h/{\rm Mpc}$ for the two-loop case (vertical lines), and applying them at lower values of $\kmaxfitcs$. We use these values as fiducial choices to calibrate the counterterms.
The relatively flat behavior as a function of $\kmaxfitcs$ and the close agreement between the gray and blue bands indicate that using counterterms fitted at higher $\kmaxfitcs$ still provides a good fit quality when applied at lower $\kmaxfitcs$. The good fit quality observed at large $\kmaxfitcs$ for the two-loop EFT does not imply that this is a regime where perturbation theory is expected to be valid \cite{Konstandin_2019}, since some degree of overfitting may be present, especially when using the power spectrum alone. The values of $\kmaxfitcs$ adopted here are chosen solely for the purpose of fitting the background counterterm evolution $\eftc_{\rm \textsc{EE2}}(z)$, shown in \reffig{c_of_z}, on top of which we allow for additional freedom. Note that the counterterms go to zero at higher redshifts, as expected. The width of the bands shows the $66$\% and $95$\% interval across the $899$ different cosmologies. The one and two-loop values of $c_{s,1}^2$ are not expected to match, since we do not subtract the cutoff-dependent $k^2P_{\rm L}$ part of $P_{1,{\rm SPT}}$ and $P_{2,{\rm SPT}}$ \cite{Baldauf:2015aha}.

After fitting those counterterms into different $\Lambda$CDM cosmologies, we use a neural network to interpolate between the different values and accelerate the calculation. However, this procedure, as defined up to this point, will not allow us to account for baryonic effects. Thus, we modify this best fit using the following procedure. The idea is to decompose the counterterms into two pieces,
\begin{equation}  \label{eq:cexp}
    \eftc(z|\Theta) = \eftc_{\rm \textsc{EE2}}(z|\Theta) + \sum_{j=1}^{n_B} \vec{c}^B_j B_{j}(z)\,,
\end{equation}
with $\eftc_{\rm \textsc{EE2}}$ obtained using the dark matter-only \textsc{EE2} best fit with cosmological parameters $\Theta$. We add a second term which expands around the dark matter-only best fit using the free parameters $c^B_j$. The redshift dependence of the correction is expressed via the cubic spline basis functions $B_{j}(z)$ as implemented in the \texttt{SciPy} library~\cite{2020SciPy-NMeth}. 
The use of a cubic spline is motivated by its piecewise-polynomial structure, which ensures smoothness while avoiding the global oscillations and boundary instabilities characteristic of high-degree polynomials. This procedure allows for a choice in the number of extra free parameters introduced in the EFT, which are set by the number of control points of the spline. 

Since we only expect the EFT counterterms to appreciably change on the order of a Hubble time, we choose slightly more conservative spline with points every $\Delta z=0.5$ from $z=0$ up to $z=4$. This fixes $n_B = 9$ in \refeq{cexp}, and thus for the two-loop EFT with five counterterms we introduce 45 free parameters, while the one-loop only introduces 9.
As we describe in \refsec{priors}, we perform a principal component analysis (PCA) in the coefficients $c^B_j$, setting the number of free parameters as the minimum number that data is sensitive to.

\begin{figure*}[t]
    \centering
    \includegraphics[width=\textwidth]{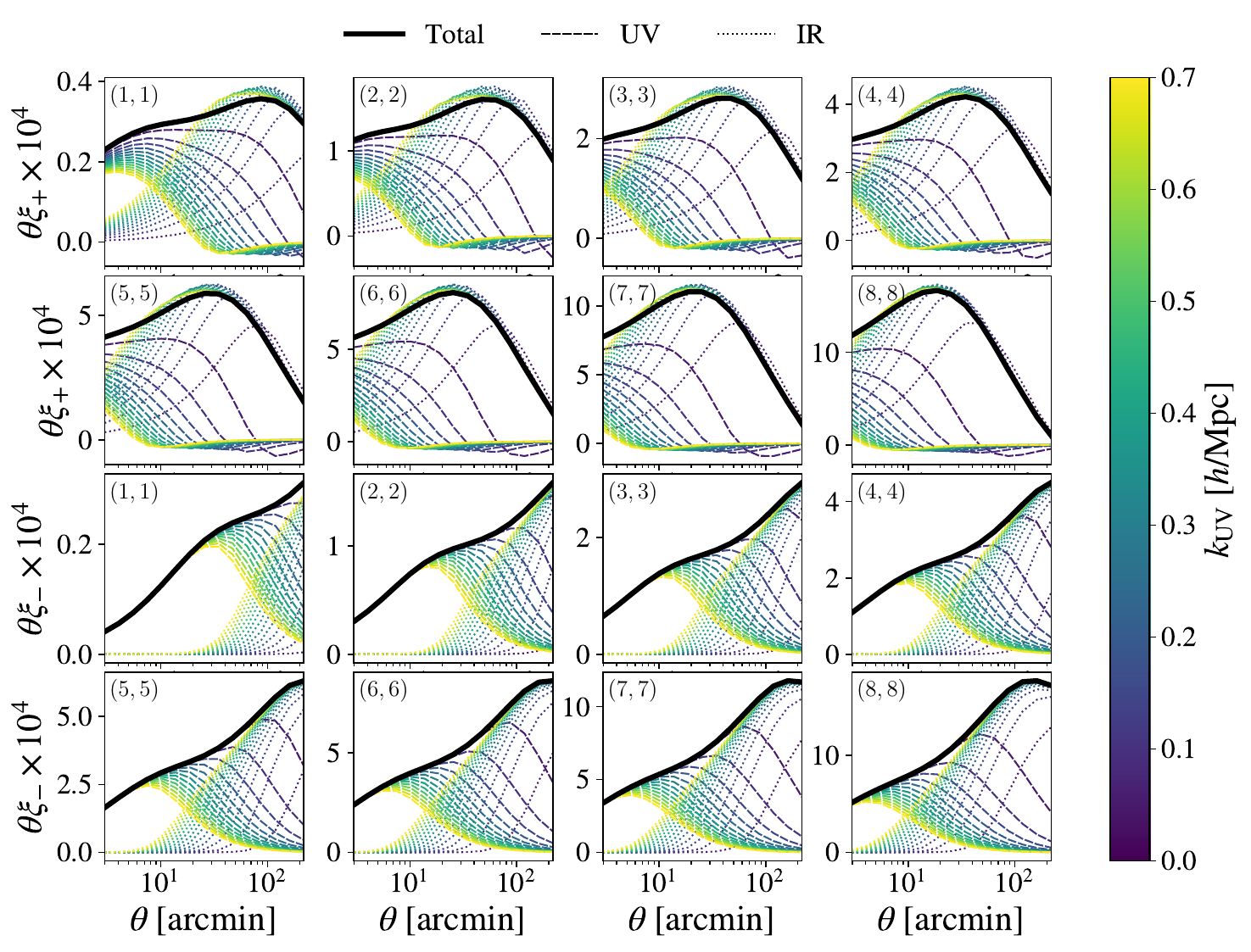}
    \caption{The UV (dashed colored) and IR (dotted colored) contributions to the total (solid black) cosmic shear for redshift bins with $i=j$, as explained in \refsec{scalecuts}. The upper eight panels display $\xi_+$ and the lower eight displays $\xi_-$. As $k_{\rm UV}$ increases, $\xi_{\pm, {\rm UV}}$ moves to smaller scales (smaller $\theta$). }
    \label{fig:xi_ir_uv}
\end{figure*}

\section{Data and analysis} \label{sec:dataanalysis}

The EFT procedure described in the previous section provides a general framework that can be applied to any lensing dataset that relies on the matter power spectrum. To test this framework, we perform forecasts for the upcoming Roman Space Telescope likelihood. In \refsec{Roman}, we describe the Roman likelihood adopted in our analysis. In \refsec{scalecuts}, we present the methodology used to determine the scale cuts and verify that $k$-modes beyond the regime of validity of the EFT are statistically insignificant.
In \refsec{priors}, we outline a method for reducing the number of counterterm parameters using principal component analysis (PCA) and derive simulation-based priors for the counterterms. Although the PCA, and therefore the priors, are survey dependent, the methodology for constructing them is fully general and can be applied to other surveys. Finally, in \refsec{Fisher}, we describe the Fisher matrix formalism used in our analysis.

\subsection{Roman Space Telescope and cosmic shear} \label{sec:Roman}

We test our analysis pipeline on Roman Space Telescope cosmic shear. As a Stage IV survey, its significantly tighter statistical uncertainties, combined with redshift coverage extending to $z=4$, provide a stringent test of the EFT modeling outlined in \refsec{EFT}. We adopt a Roman configuration with eight tomographic redshift bins extending to $z=4$ shown in Fig.~\ref{fig:roman_nz}. 

Assuming a spatially flat Universe and adopting the Limber approximation \cite{Limber}, which is justified given that we restrict the analysis to angular scales $\theta \lesssim 250$ arcminutes, the tomographic lensing power spectrum is given by 
\begin{equation} \label{eq:Cell}
C_{\ell}^{ij} = \int_{0}^{\chi_H} d\chi 
\frac{q^i(\chi) q^j(\chi)}{\chi^2}
P_{\rm NL}\left( k = \frac{\ell + 1/2}{\chi}, \chi \right),
\end{equation}
where $\chi$ is the comoving radial distance, $\chi_H$ is the comoving horizon distance, and $P_{\rm NL}$ denotes the non-linear matter power spectrum. The lensing efficiency for tomographic bin $i$ is defined as~\cite{Doux_2021,krause2021darkenergysurveyyear,Friedrich_2021}
\begin{equation} \label{eq:lenseff}
q^i(\chi) =
\frac{3}{2}
\frac{\Omega_{\rm m} H_0^2}{c^2}
\frac{\chi}{a(\chi)}
\int_\chi^{\chi_H} d\chi'
n^i(\chi')
\frac{\chi' - \chi}{\chi'} \,,
\end{equation}
where $\Omega_{\rm m}$ is the matter density today, $H_0$ is the Hubble constant, $c$ is the speed of light, $a(\chi)$ is the scale factor, and $n^i(\chi)$ is the normalized redshift distribution of source galaxies in bin $i$ (see Fig.~\ref{fig:roman_nz}).

Different analyses adopt different prescriptions for the non-linear matter power spectrum. For example, HSC~\cite{Dalal:2023olq}, KiDS~\cite{Wright:2025xka} and DES \cite{DES:2026mkc} use \texttt{HMCode} \cite{Mead:2020vgs}, with baryonic corrections calibrated on the \texttt{BAHAMAS} hydrodynamical simulations \cite{McCarthy:2016mry}. DES fixes the AGN feedback temperature parameter, while KiDS and HSC marginalize over it.
In contrast, the goal of this work is to avoid phenomenological baryonic modeling altogether and instead employ the one- and two-loop EFT description of the matter power spectrum outlined in \refsec{reviewEFT}, as already considered at one loop by \cite{damico2025cosmologicalanalysisdes3times2pt}.

The cosmic shear two-point correlation functions in an angular bin $[\theta_{\rm min}, \theta_{\rm max}]$ are given by~\cite{krause2021darkenergysurveyyear,Friedrich_2021,Xu:2025evn}
\begin{equation} \label{eq:shear}
\xi_{\pm}^{ij}(\theta) =
\sum_{\ell}
\frac{2\ell+1}{2\pi \ell^2 (\ell+1)^2}
\left[
\overline{G_{\ell,2}^{+}(\cos\theta)
\pm
G_{\ell,2}^{-}(\cos\theta)}
\right]
C^{ij}_\ell \,,
\end{equation}
where $G^{\pm}_{\ell,2}(\cos\theta)$ are geometric functions related to the associated Legendre polynomials, as given in~\cite{stebbins1996weaklensingcelestialsphere}. The bar denotes averaging over the finite angular bins. In this analysis, the cosmic shear two-point functions are evaluated in 15 logarithmically spaced angular bins spanning $2.5$ to $250$ arcminutes.

The Roman likelihood is a Gaussian likelihood on a vector consisting of the concatenated cosmic shear correlation functions with log-likelihood given by (up to an additive normalization constant)
\begin{equation}\label{eq:likelihood}
    \log \mathcal{L} \simeq -\frac{1}{2} \left[ \boldsymbol{m}(\Theta) - \boldsymbol{d}\right]^{T} \Sigma^{-1} \left[ \boldsymbol{m}(\Theta) - \boldsymbol{d} \right] = -\frac{1}{2}\chi^2 \,,
\end{equation}
where $\boldsymbol{m}(\Theta)$ is the datavector generated with cosmological parameters $\Theta$, $\boldsymbol{d}$ is the synthetic cosmic shear datavector we use for the analysis and $\Sigma$ the covariance matrix computed using \textsc{CosmoCov}~\cite{2017MNRAS.470.2100K,Fang__2020}.

We modified the Roman likelihood, implemented in \textsc{CosmoLike}~\cite{2017MNRAS.470.2100K}, to use the EFT non-linear power spectrum and sample the counterterm amplitudes $c^B_j$ (or principal components of this counterterms hyperspace, as described in \refsec{priors}) from \refeq{cexp}. We emphasize again that only the counterterm background piece $\eftc_{\rm \textsc{EE2}}(z|\Theta)$ is fit minimizing the $\chi^2_{\rm MPS}$ metric from \refeq{chi2MPS}, based on a set of \textsc{EE2} emulated power spectra, while deviations from this background part are to be determined (or marginalized over) by data.

We consider two scenarios for $\boldsymbol{d}$ to test our EFT implementation. First, we forecast $S_8 = \sigma_8\sqrt{\Omega_m/0.3}$ for a noiseless, dark matter only datavector generated using \textsc{EE2} using the cosmological and nuisance parameters in \reftab{fiducial_cosmo}. The EFT counterterms, being calibrated to a dark matter-only power spectrum as described in \refsec{eft_evolution}, is expected to fit the data with few additional parameters, allowing us to see the range of applicability of each EFT loop order. The second scenario forecasts $S_8$ for noiseless datavector contaminated with either \textsc{BAHAMAS}~\cite{McCarthy:2016mry,McCarthy_2018} or \textsc{OWLS-AGN}~\cite{Le_Brun_2014} hydrodynamic simulations, the latter presenting stronger baryonic feedback than  the former, providing a stringent test of our modelling of the counterterms. The simulations provide the amount of suppression of the dark matter only power spectrum (which is still computed with \textsc{EE2}) resulting from baryonic physics on small scales.
 
\begin{table}
	\centering
	\renewcommand{\arraystretch}{1.3}
	\begin{tabular}{lcc}
		Parameter & Fiducial Value & Prior \\
		\hline
		\emph{Cosmological Parameters} & & \\
		$A_s 10^{9}$ & $2.08$ & $[1.7,2.5]$ \\
		$n_s$ & $0.959$ & Fixed \\ 
		$H_0$ & $66.2$ & Fixed \\ 
		$\Omega_{\rm b}$ & $0.0504$ & Fixed \\ 
		$\Omega_{\rm m}$ & $0.329$ & $[0.24,0.40]$ \\
		\hline
		\emph{Source Photo-}$z$ & & \\
		$\Delta z_{S,i}$ for $i=1,2,\ldots 8$ & $0.0$ & $\mathcal{N}(0.0,0.002)$ \\
		\hline
		\emph{Intrinsic Alignment} & \\
		$a_1$ & $0.6$ & $[-5.0,5.0]$ \\
		$\eta_1$ & $-1.5$ & $[-5.0,5.0]$\\
		\hline
		\emph{Shear Calibration} & \\
		$ m^i$ for $i=1,2,\dots 8$ & $0.0$ & $\mathcal{N}(0.0,0.005)$ \\
		\hline
	\end{tabular} 
	\caption{The fiducial cosmological parameters and priors adopted in this manuscript. These parameters correspond to the maximum-a-posteriori of the CMB first peak prior.
    We use these cosmological parameters to generate the synthetic Roman cosmic shear datavectors used throughout this manuscript. We also show the photometric redshift uncertainty, intrinsic alignment, and shear calibration nuisance parameters. The priors are used to get a Fisher-estimated error for $S_8$, as explained in \refsec{Fisher}. }
	\label{tab:fiducial_cosmo}
\end{table}

Another key aspect of shear–shear correlation analyses is the modeling of intrinsic alignments (IA) \cite{Hirata:2004gc}. Intrinsic alignments arise when galaxy shapes become correlated with the surrounding large-scale tidal field during their formation and evolution. Although this effect is independent of gravitational lensing, it can mimic or contaminate the lensing signal if not properly modeled. We adopt the non-linear alignment (\textsc{NLA}) model for intrinsic alignments \cite{Bridle:2007ft,krause2021darkenergysurveyyear}, in which the IA contribution is described by a single amplitude parameter $a_1$, together with a redshift scaling of the form $(1+z)^{\eta_1}$. We therefore do not consider the more complete EFT-based descriptions of galaxy shapes \cite{Vlah:2019byq,Vlah:2020ovg,Chen:2023yyb} or the tidal alignment and tidal torquing (\textsc{TATT}) model~\cite{Blazek_2019}. The impact of using these models for Roman cosmic shear warrants an independent investigation.

We also marginalize over two more types of nuisance parameters important for cosmic shear. The photometric redshift uncertainty $\Delta z_{S,i}$ for $i\in\{1,\ldots,8\}$ assigns an uncertainty in the redshift distribution of source galaxies for each galaxy sample $i$. This amounts to the transformation $n^{i}(z) \rightarrow n^{i}(z+\Delta z_{S,i})$ in equation \refeq{lenseff}. The shear calibration bias $m_{i}$ for $i\in\{1,\ldots,8\}$ is applied directly to the cosmic shear correlation function by the transformation $\xi_{\pm}^{ij} \rightarrow (1+m^i)(1+m^j) \xi_{\pm}^{ij}$. 

\subsection{IR-UV decomposition and scale cuts} \label{sec:scalecuts}

In line with the EFT methodology, we aim to restrict our analysis to Fourier modes up to a maximum scale $\kmax$ (the IR regime), where the perturbative expansion at a fixed loop order is expected to be reliable for a given survey precision. Smaller-scale modes (the UV regime) are effectively marginalized over through the inclusion of counterterms.\footnote{For galaxy clustering analyses, the standard procedure is to adopt $\kmaxfitcs = \kmax$ (often denoted $k_{\rm max}$) and marginalize over a set of counterterms defined independently for each redshift bin and tracer sample. In lensing analysis, however, we integrate over redshift along the line of sight. We therefore parametrize the redshift evolution of the counterterms as described in \refsec{eft_evolution}, using the background $\eftc_{\rm \textsc{EE2}}$ fit evaluated at $\kmaxfitcs$, while restricting our analysis to $\kmax \leq \kmaxfitcs$.}
In three-dimensional power spectrum analyses, the choice of $\kmax$ is typically determined by scanning over progressively smaller scales until the EFT description fails according to a specified criterion, which can be a significant increase in $\chi^2$, the appearance of parameter biases, or the growing importance of higher-loop contributions. In weak lensing, however, the lensing kernel integrates over a broad range of redshifts. As a result, contributions from IR and UV modes become mixed in the projected observables.

\begin{figure}[t]
    \centering
    \includegraphics[width=\columnwidth]{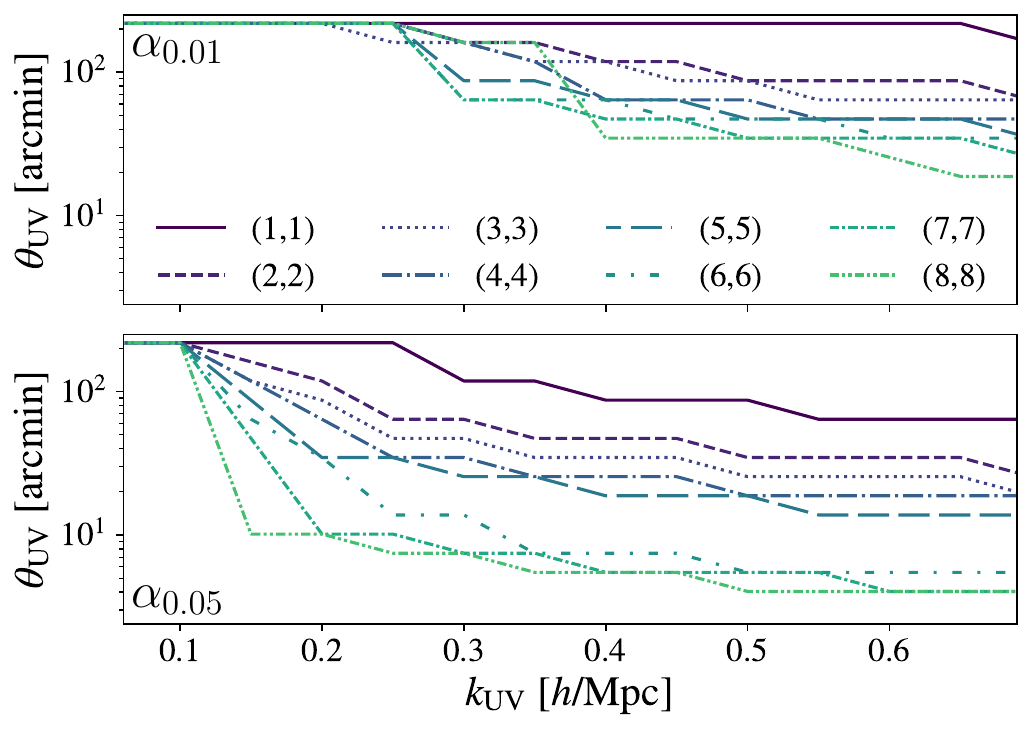}
    \caption{The adopted scalecuts for the shear-shear correlations, as outlined in \refsec{scalecuts}. We only display the cuts for the same galaxy sample redshift bins $(i=j)$. The top panel corresponds to the UV-IR split with $\alpha=0.01$, while we use $\alpha=0.05$ for the bottom panel. The conservative $\alpha_{0.01}$ scale cuts do not have any unmasked data until $\kmax=0.25$ $h$/Mpc, while the more aggressive choice allows one to start at $\kmax=0.15$ $h$/Mpc. Additionally, $\alpha_{0.01}$ always masks more data than the $\alpha_{0.05}$, and both are monotonic in $\kmax$.}
    \label{fig:scalecuts}
\end{figure}

The problem of defining scale cuts therefore translates into establishing a mapping $\thetamin(\kmax)$, where $\thetamin$ denotes the minimum angular scale kept in the cosmic shear analysis. The purpose of this mapping is to ensure that the observables are insensitive to contributions from modes with $k > \kmax$. 
In this work, we follow \citet{derose2025lensingcounternarrativeeffective}, that aimed to find angular scale cuts such that the residual sensitivity to UV modes has a negligible impact on the inference, consistent with our goal of implementing a conservative and self-consistent EFT treatment. Based on \citet{derose2025lensingcounternarrativeeffective}, we decompose the non-linear matter power spectrum in Eq.~\eqref{eq:Cell} into IR and UV components,
\begin{align}
\label{eq:cl_mode_split}
C_{\ell}^{ij} =& \int_{0}^{\chi_H} d\chi \frac{q^i(\chi)q^{j}(\chi)}{\chi^2} \\
&\times \left[ 
    P_{\rm NL}\left(\frac{\ell+1/2}{\chi},\chi \right)
    \Theta\bigg(\kmax/D(\chi)-\frac{\ell+1/2}{\chi}\bigg) \right. \notag\\
&\quad+ \left.
    P_{\rm NL}\left(\frac{\ell+1/2}{\chi},\chi \right)
    \Theta\bigg(\frac{\ell+1/2}{\chi}-\kmax/D(\chi)\bigg)
\right] \notag\\
\equiv& \,C_{\ell,{\rm IR}}^{ij} + C_{\ell,{\rm UV}}^{ij}\,, \notag
\end{align}
where $\Theta(\cdot)$ denotes the Heaviside step function, which explicitly separates contributions from modes satisfying $k \leq \kmax/D(\chi)$ and $k > \kmax/D(\chi)$. This decomposition propagates directly to the shear correlation functions $\xi_{\pm}$, \refeq{shear}. In \reffig{xi_ir_uv}, we illustrate the IR and UV contributions for different values of $\kmax$, using the \textsc{EE2} non-linear matter power spectrum for $P_{\rm NL}$. For a fixed angular scale $\theta$, the IR contribution (solid colored lines) decreases as $\kmax$ is lowered, while the UV contribution (dashed lines) correspondingly increases, relative to the total signal (black solid line). Once more, we assume a $\kmax/D(\chi)$ scaling of the non-linear scale. A more detailed analysis on how the non-linear scale evolve as a function of redshift could in principle optimize the scalecuts.

Therefore, for any fixed multipole $\ell$ or angular scale $\theta$, both IR and UV $k$-modes contribute to the total signal such that one cannot simply discard the UV contribution. Instead, as also suggested by \cite{Braganca:2020nhv}, we model modes with $k>\kmax$ using \textsc{EE2} (alternatively, \texttt{Halofit} \cite{Takahashi:2012em}) and impose continuity at the matching scale by setting
$P(k>k_{\rm UV}/D(z)) = AP_{\rm EE2}$ with $A$ an amplitude such that $P_{\rm EFT}(k = k_{\rm UV}/D(z)) = A P_{\rm EE2}(k = k_{\rm UV}/D(z))$.  
To still remain insensitive to the UV signal, we require the total contribution of the UV modes to the total cosmic shear signal to be smaller than some fraction $\alpha$, which reflects the statistical precision of the survey.\footnote{This definition of the UV contamination depends on what is assumed for the UV completion, but should be relatively stable as long as there is no sharp increase in the power spectra on these scales.} We find $\thetamin$ that guarantees that every angular bin we include in our analysis obeys
\begin{equation} \label{eq:criteriacut}
    \frac{\left\rvert\xi^{ij}_{\rm UV}(\theta)\right\rvert}{\left\rvert\xi^{ij}_{\rm UV}(\theta)\right\rvert + \left\rvert\xi^{ij}_{\rm IR}(\theta)\right\rvert} < \alpha\,.
\end{equation}
For reference, \cite{derose2025lensingcounternarrativeeffective} adopted a $5\%$ UV convergence criterion for DES and $1\%$ for LSST-Y10, while \cite{Doux_2021} imposed a $5\%$ upper limit on the contribution from $k>\kmax$ using a response-function approach. 

In this work, we consider both $\alpha=0.05$ (denoted $\alpha_{0.05}$), corresponding to a more aggressive choice, and $\alpha=0.01$ (denoted $\alpha_{0.01}$), corresponding to a more conservative amount of UV contamination.
 We further note that the criteria \refeq{criteriacut} does not strictly guarantee monotonicity of the $\thetamin(\kmax)$ relation (i.e., larger $\kmax$ necessarily leading to smaller $\thetamin$), since $\xi_+^{ij}$ presents a zero crossing, as seen in figure \reffig{xi_ir_uv}. We then enforce monotonicity by certifying that if a $\theta$ bin is unmasked for a given $\kmax$, then it is also unmasked for smaller values of $\kmax$. Additionally, we ensure that there are not any sharp changes in $\theta_{\rm UV}$ for small changes in $\kmax$.

\begin{figure}[t]
    \centering
    \includegraphics[width=\columnwidth]{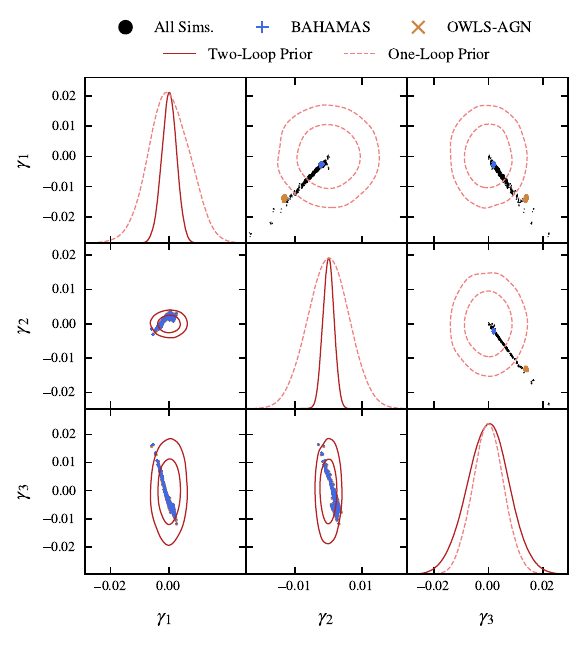}
    \caption{The prior on the first three (PCA-projected) EFT parameters $\gamma_i$ for $\alpha_{0.05}$ and $\kmax = 0.5\,h/$Mpc. The lower triangle shows the two-loop prior, and the upper triangle shows the one-loop prior. The black points correspond to the baryon-contaminated counterterms (see \refsec{priors}), measured across $899$ cosmologies, projected onto the amplitude of PCA basis $\gamma_i$. The yellow $\times$ are PCA amplitudes $\gamma_i$ for the \textsc{OWLS-AGN} simulation with $\log_{10}[T_{\rm AGN}/(1\,\text{K})] = 8.0$, and the blue $+$ are the PCA amplitudes for the \textsc{BAHAMAS} simulation with $\log_{10}[T_{\rm AGN}/(1\,\text{K})] = 7.8$. In the two-loop prior, the two simulations overlap each other. The priors contain the simulations to within roughly two standard deviations centered at $\gamma_i=0$ and neglecting the cross-correlations between the amplitudes. }
    \label{fig:priors}
\end{figure}

The resulting $\thetamin(\kmax)$ relations for $\alpha_{0.01}$ and $\alpha_{0.05}$ are shown in Fig.~\ref{fig:scalecuts}. As expected, the more conservative choice $\alpha_{0.01}$ removes a larger number of data points. Different tomographic bins are shown in distinct colors. Note that lower-redshift bins, which probe more non-linear scales, require more stringent cuts (larger $\thetamin$).

\subsection{Principal component analysis and priors}\label{sec:priors}


While our procedure introduces $45$ additional nuisance parameters at two loops and $9$ additional nuisance parameters at one-loop for $n_B = 9$ in \refeq{cexp}, the data is unlikely to be sensitive to all them. For example, the lensing efficiency defined in \refeq{lenseff} vanishes at $z=0$ and at the redshift of the source galaxy $z_s$, meaning the data is insensitive to the value of $c^B_j$ at $z=0$ and $z=z_s$. 

One way to identify meaningful parameters is using a principal component analysis (PCA), which organizes the parameters into linear combinations ordered by how well they are constrained by the likelihood. In practice, this is done by concatenating {\it all} the 45 (9) two(one)-loop counterterms from \refeq{cexp} into a vector, diagonalizing the Fisher matrix, and ordering its eigenvectors in decreasing eigenvalues. Since the PCA depends on the scale cuts, we perform this PCA at every $\kmax$ and for both $\alpha_{0.01}$ and $\alpha_{0.05}$. The PCA, therefore, defines a new set of parameters $\gamma_i$, with the index running over the ranked principal components, to be fitted by data. We present the first few components for each of the one and two-loop counterterms in \refapp{PCA}.
We discuss the effect of the number of principal components $N_{\rm PCA}$ on the $\chi^2$ in \refsec{results}, where the results on the inferred $S_8$ are also considered. 

Although the PCA informs which counterterms lensing constrains the most, the data alone may still allow for unphysical values of the PCA amplitudes. A reasonable prior can further constrain the counterterms to follow physically motivated values. Different kinds of priors have been adopted in the literature. For example, \cite{damico2025cosmologicalanalysisdes3times2pt} adopts $\mathcal{O}(1)$ theoretically-motivated from the EFT perspective priors while \cite{Barreira:2020kvh, Ivanov:2024hgq, Ivanov:2024xgb, Akitsu:2024lyt, Shiferaw:2024ehr, ivanov2025simulationbasedpriorssimulationsanalytic} adopt simulation-based priors.

In this work, we adopt a simulation-based approach using baryonic feedback data to ensure we have well motivated priors.
To calibrate the priors of the EFT, we use baryonic feedback data from: 
\textsc{BAHAMAS}~\cite{McCarthy_2018},
\textsc{Antilles}~\cite{Salcido_2023},
\textsc{OWLs}~\cite{Le_Brun_2014},
\textsc{Illustris}~\cite{Genel_2014,Vogelsberger_2014} and 
\textsc{IllustrisTNG-100}~\cite{Pillepich_2017,Springel_2017,Naiman_2018,Marinacci_2018,Nelson_2017},
\textsc{Eagle}~\cite{Crain_2015,Schaye_2014},
\textsc{MassiveBlack2}~\cite{Khandai_2015},
and \textsc{HorizonAGN}~\cite{Chisari_2018}. 
We first fit the counterterms to the \textsc{EE2} power spectrum contaminated by each of the baryonic feedback simulations at each of the $899$ cosmological parameters considered and at the splined redshifts, which is $z=[0.0, 0.5,\ldots,4.0]$ in our application. When performing the fits, we assume that the baryonic feedback correction to the non-linear power spectrum is independent of cosmological parameters. We then construct a prior on the PCA amplitudes $\gamma_i$ projecting the baryon-valued counterterms onto the PCA basis and performing a Gaussian approximation on the baryon simulation amplitudes. To make sure the prior is conservative, we force the mean to be zero, eliminate all correlations between the amplitudes, and scale the variances by a factor five. These changes ensure that most simulations we consider, including the ones which are excluded by lensing data~\cite{Xu:2025evn}, are within a few standard deviations of the prior. 

An example of the prior adopted for the $\alpha_{0.05}$ coefficients at $\kmax = 0.5\,h/{\rm Mpc}$ is shown in \reffig{priors}. As emphasized above, the priors are recalibrated at \emph{each} value of $\kmax$ and independently for both $\alpha_{0.01}$ and $\alpha_{0.05}$. The contours display the one- and two-loop priors projected onto the amplitudes of the first three eigenfunctions, $\gamma_1$, $\gamma_2$, and $\gamma_3$. At $\kmax=0.5\,h/\rm{Mpc}$, the prior in the two-loop terms are narrower than for the one-loop.
Note the relatively small magnitude of these amplitudes compared to the typical value of $\eftc_{\rm \textsc{EE2}}$ in \refeq{cexp} (see \reffig{c_of_z}), indicating that the background-calibrated counterterms, which parametrize dark matter non-linearities, are generically larger than the baryon-induced corrections captured by the $\gamma_i$.

Simulation results are overlaid as black points in \reffig{priors}, with OWLS-AGN and BAHAMAS highlighted separately yellow and blue, respectively. 
In the two-loop prior, the two simulations overlap each other, suggesting that the part of the priors that is given by the baryonic feedback is subleading for the two-loop compared to other contributions that can be absorbed by $\gamma_i$. Those can be either corrections due dark matter non-linearities not absorbed by $\eftc_{\rm \textsc{EE2}}(z|\Theta)$ fit in \refeq{cexp} or also residual interpolation errors from the fit of $\eftc_{\rm \textsc{EE2}}$ in $\Theta$-space.
For $\alpha_{0.05}$ at $\kmax = 0.5\,h/{\rm Mpc}$, the OWLS-AGN model lies outside the one-loop prior region, even after enlarging the priors by a factor five to be conservative. That reflects its comparatively strong feedback implementation relative to the other baryonic scenarios considered. This does not compromise our analysis since we obtain consistent results for the one-loop EFT fitted to OWLS-AGN and BAHAMAS, as discussed in \refsec{baryons}. A systematic study of how different baryonic feedback prescriptions are absorbed into the EFT counterterms is deferred to future work.

The priors we use on the cosmological, photometric redshift bias, intrinsic alignment, and shear calibration bias parameters are shown in table \reftab{fiducial_cosmo}. We also include an additional Gaussian prior on the cosmological parameters. We follow~\citet{Zhong_2023} by including the first acoustic peak ($35<\ell<396$ along with $\ell<30$ for the CMB EE power spectrum) of the CMB as a prior on $\Omega_{\rm m} h$, effectively reducing the allowed $k_{\rm NL}$ range and not allowing for cosmologies with completely distinct perturbative behavior. Furthermore, the first peak alone is only informative on $n_s$, $H_0$ and $\Omega_{\rm b}$, which are mostly unconstrained by cosmic shear alone. Also, the smoothing of acoustic peaks from gravitational lensing does not greatly affect parameter constraints on its own.

\subsection{Fisher forecast and parameter bias estimation} \label{sec:Fisher}

\begin{figure*}[t]
    \centering
     \includegraphics[width=\textwidth]{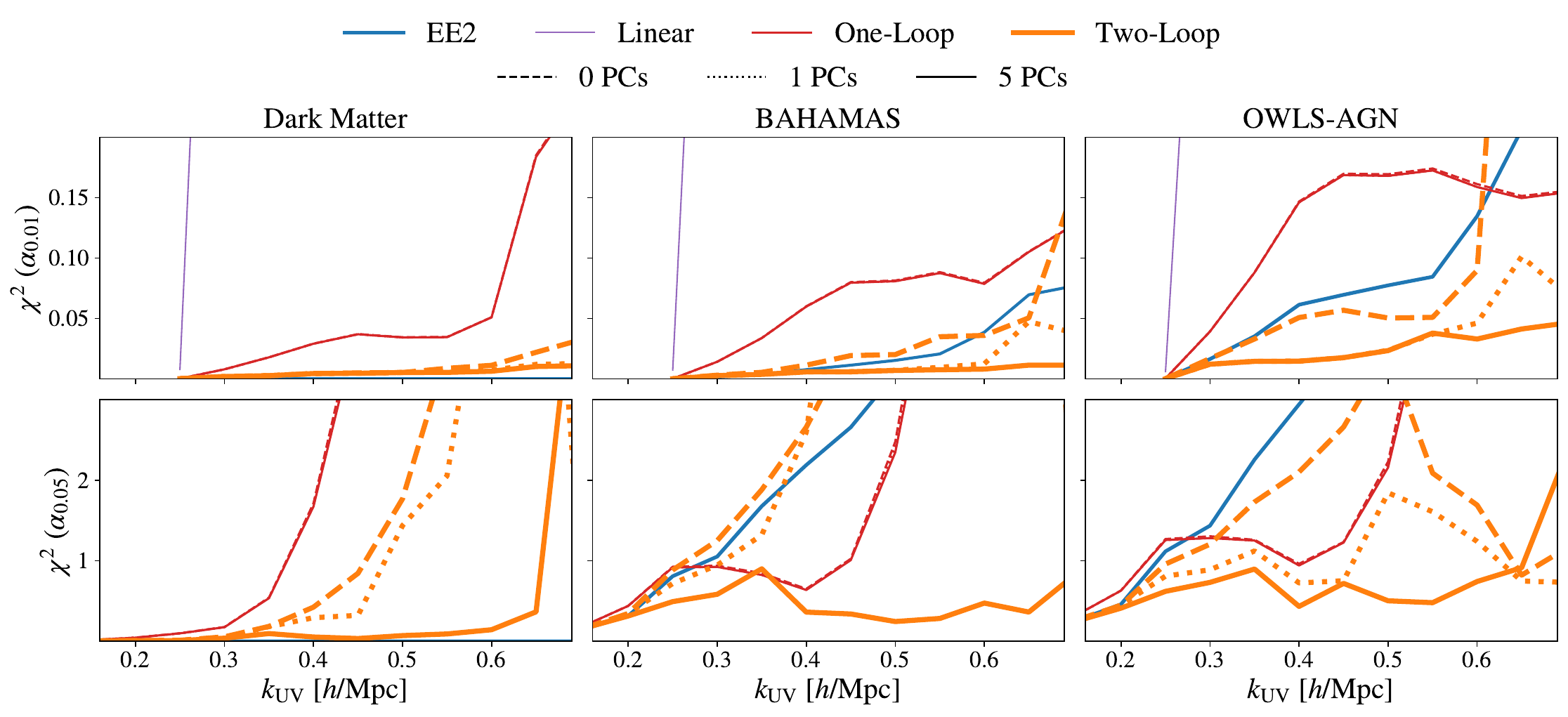}
    \caption{The $\chi^2$ of the Roman likelihood as a function of $\kmax$, as defined in \refeq{likelihood}, for the dark matter-only case (left), BAHAMAS contaminated (middle) and OWLS-AGN contaminated (right). The upper panel considers $1\%$ of UV contamination ($\alpha_{0.01}$) and the lower panel $5\%$ ($\alpha_{0.05}$). We show results for \textsc{EE2} (blue, with $\chi^2=0$ by construction for the dark matter-only case), linear (purple), one (red) and two-loop (orange) EFT modelling of the matter power spectrum. For the EFT modelling we consider different values of principal components (PCs) for the counterterms. One-loop results with different PCs overlap.}
    \label{fig:chi2}
\end{figure*}

We perform a Fisher forecast around the fiducial cosmology specified in Table~\ref{tab:fiducial_cosmo}.
We estimate the shift in the maximum a posteriori (MAP) parameters induced by model mismatches defining the parameter displacement
\begin{equation}
    \boldsymbol{\delta} \equiv \Theta_{\rm MAP} - \Theta_{\rm fid}\,.
\end{equation}
The linearized solution for the MAP shift is
\begin{equation}
    \boldsymbol{\delta}
    =
    F_{\mathcal{P}}^{-1}
    \left[
        J^{T}
        \Sigma^{-1}
        \bigl(\boldsymbol{d} - \boldsymbol{m}_{\rm fid}\bigr)
        -
        F_{\Pi}(\Theta_{\rm fid} - \boldsymbol{\mu})
    \right],
\end{equation}
where we define
\begin{equation}
    J \equiv 
    \left.
    \frac{\partial \boldsymbol{m}(\Theta)}{\partial \Theta}
    \right|_{\Theta = \Theta_{\rm fid}},
    \qquad
    \boldsymbol{m}_{\rm fid} \equiv 
    \boldsymbol{m}(\Theta_{\rm fid}).
\end{equation}
Here $\Sigma$ is the data covariance matrix and 
\begin{equation}
    F_{\mathcal{P}} = F + F_{\Pi}\,,
\end{equation}
is the posterior Fisher matrix, given by the sum of the likelihood Fisher matrix
\begin{equation}
    F = J^{T}\Sigma^{-1}J\,,
\end{equation}
and the prior Fisher matrix $F_{\Pi}$. The vector $\boldsymbol{\mu}$ denotes the center of the Gaussian prior (derived from the first acoustic peak of the CMB).

Cosmological parameters generically become biased when variations in cosmology partially mimic the impact of baryonic effects in the datavector and no other nuisance parameters, such as counterterms, can absorb the baryonic physics. In this case, the parameter displacement corresponds to the projection of the baryonic modeling error onto the cosmological parameter directions. If none of the parameters mimic baryons, then no parameter shift occurs, and only the goodness of fit to the data is affected.

To get the final constraint on $S_8$, we run an MCMC on the Fisher-approximated likelihood with the Gaussian prior derived from the first acoustic peak of the CMB and the priors outlined in table \reftab{fiducial_cosmo}. We fix the parameters $n_s$, $H_0$, and $\Omega_{\rm b}$, which are prior dominated when only cosmic shear is used. We use the approximation formula from~\citet{Secco_2023} to determine $S_8$, which works to about a $1\%$ accuracy within $10\sigma$ of the Planck 2018 best fit cosmological parameters~\cite{Planck:2018vyg}.

\section{Results} \label{sec:results}

In this section we present the forecasted $S_8$ constraints for Roman, while fixing the other cosmological parameters. We consider two scenarios, as described in \refsec{Roman}. First, in \refsec{darkmatter}, a dark matter-only datavector generated by \textsc{EE2} and second, in \refsec{baryons}, a baryon-contaminated data. 

\subsection{Dark matter only}\label{sec:darkmatter}

\begin{figure*}[t]
    \centering
    \includegraphics[width=\textwidth]{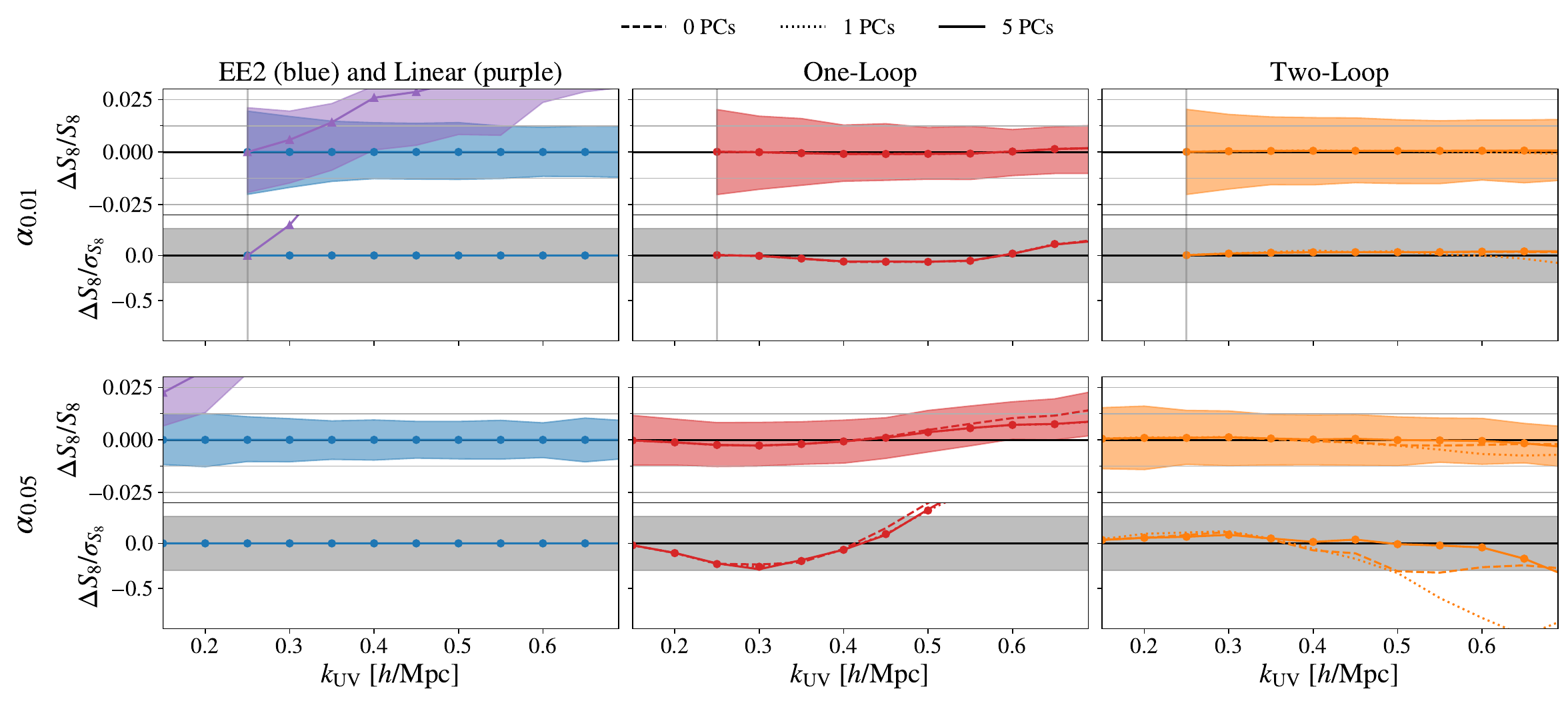}
    \caption{The forecasted $S_8$ constraints as a function of $\kmax$ for a noiseless dark matter-only datavector computed using \textsc{EE2} for both the $\alpha_{0.01}$ (first row) and $\alpha_{0.05}$ (second row) scale cuts. We show results for linear modelling and \textsc{EE2} (purple and blue respectively) in the left panels, the latter having vanishing bias by construction. For low $\kmax$ and $\alpha_{0.01}$ the data are completely masked, as described in the top panel of \reffig{scalecuts}. One-loop and two-loop EFT results are shown in the middle and right panels, with different line styles corresponding to different values of principal components (PCs) used for the counterterm. The bottom panels of each figure indicate the bias in the parameters, with the gray band corresponding to $0.3\sigma$. The colored band corresponds to the $1\sigma$ estimated error regions, displayed for the 5(1)-component case for the two(one)-loop EFT.
    }
    \label{fig:s8_dm}
\end{figure*}
\begin{figure*}[t]
    \centering
     \includegraphics[width=\textwidth]{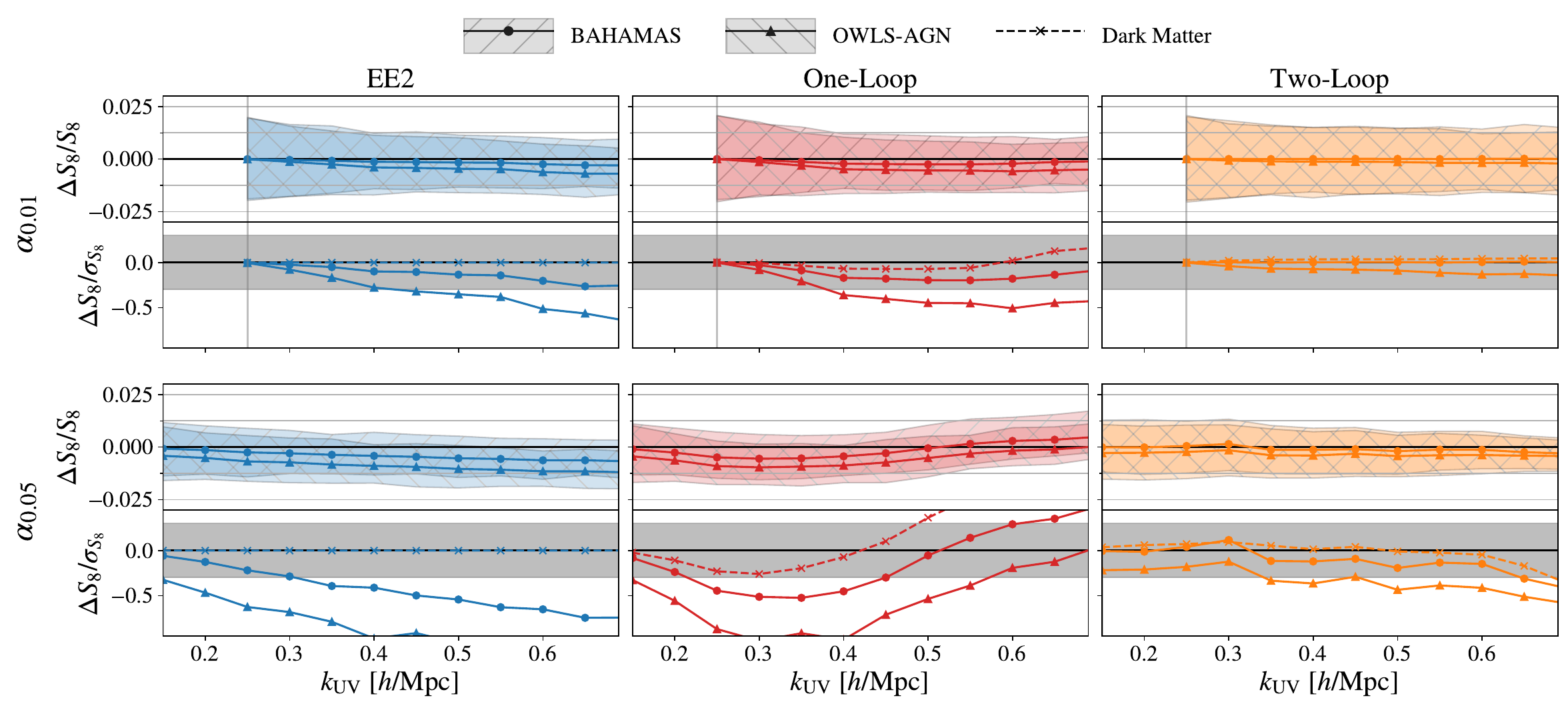}
    \caption{The forecasted $1\sigma$ constraints in $S_8$ for noiseless synthetic datavectors contaminated with the \textsc{BAHAMAS} (dot markers and the '$/$' hatching) and the \textsc{OWLS-AGN} (triangle markers and '\textbackslash' hatching) baryonic feedback corrections. In the one-loop case, we show that constraints with $N_{\rm PCA}=1$, while for the two loop we show constraints with $N_{\rm PCA}=5$.  For reference, we show the $S_8$ bias for the dark matter case from \reffig{s8_dm} with the same $N_{\rm PCA}$ in the dashed line with '$\times$' markers.    }
    \label{fig:s8_baryons}
\end{figure*}

As a validation test of our EFT implementation, we first present results for a noiseless, dark matter–only datavector generated with \textsc{EE2}. This setup provides an idealized scenario to test the constraining power of the one- and two-loop EFT predictions. When analysing the results, it is important to have in mind that the background values of the counterterms were calibrated using dark matter-only spectra, following \refeq{cexp}.

The left panel of \reffig{chi2} shows the value of $\chi^2$ for the dark matter-only scenario, both for $\alpha_{0.01}$ (top row) and $\alpha_{0.05}$ (bottom row). By construction, $\chi^2 = 0$ for \textsc{EE2} for the dark matter-only case. The linear scenario almost never produces a good fit for the scales considered, not even appearing in the plot for $\alpha_{0.05}$, reflecting the fact that linear theory is reliable only at very large scales. For the most conservative UV contamination scenario $\alpha_{0.01}$, both one and two-loop provide $\chi^2 \ll 1$, with the one-loop starting to see an increase in $\chi^2$ only at $\kmax = 0.6\,h/\mathrm{Mpc}$.\footnote{We do not compare the values of $\kmax$ reported here with the expected one-loop galaxy-clustering three-dimensional validity range $k_{\rm UV} \lesssim 0.2\,h/\mathrm{Mpc}$ at $z=0$, due to the integral in the line-of-sight and with the scale cuts being actually in $\thetamin$, as described in \refsec{scalecuts}.} We find that including additional principal components (PCs), indicated by the different line styles, yields negligible improvement in the one-loop case. This is expected, since the background counterterm values were already calibrated to \textsc{EE2}, as described in \refsec{eft_evolution}. 

When considering $\alpha_{0.05}$, the one-loop provides a good fit up to $\kmax = 0.35\,h/\mathrm{Mpc}$ for any number of PCs, while the two-loop case is very sensitive to the number of PCs considered for the counterterms. The improvement in $\chi^2$ with the number of PCs may come from two different reasons. First, it may add extra non-linear corrections and fix interpolation errors on top of the background fit $\eftc_{\rm \textsc{EE2}}(z|\Theta)$ in \refeq{cexp}. Second,  given that the counterterm background values were already calibrated using \refeq{cexp}, this improvement suggests that the PCs are likely partially overfitting data beyond $\kmax = 0.5\,h/\mathrm{Mpc}$. 
Overfitting is a known concern for two-loop EFT analyses of the two-point function \cite{Konstandin_2019} and can revealed by including higher-order $n$-point functions \cite{Baldauf:2021zlt}. In our implementation, overfitting in the counterterms can occur between either term in right-hand side of \refeq{cexp}, with the first term manifesting as a shift in cosmological parameters.

The upper panels of \reffig{s8_dm} show the forecasted constraint on $S_8$ as a function of $k_{\rm UV}$ for the values of UV contamination $\alpha_{0.01}$ (top row) and $\alpha_{0.05}$ (bottom row). The lower panels in each row display the corresponding bias expressed in units of $\sigma$, where the $0.3\sigma$ threshold commonly adopted by DES~\cite{Amon_2022,Secco_2022,krause2021darkenergysurveyyear} is indicated by a gray band. Results are shown for \textsc{EE2} and linear modeling (left column), one-loop EFT (middle column), and two-loop EFT (right column). 
By construction, \textsc{EE2} (blue) exactly reproduces the fiducial datavector and therefore exhibits no bias. In contrast, the linear model (purple) is strongly biased in nearly all cases and exhibits a bias that increases monotonically with $\kmax$.

The situation improves substantially at one loop. The middle panels of \reffig{s8_dm} show the results for the one-loop case for different values of principal components considered, which almost perfectly overlap. We show the error (shaded band) for the case with one PC since it allows some freedom to absorb contributions not captured by the background term $\eftc_{\textsc{EE2}}(z|\Theta)$ in \refeq{cexp}, even though it does not affect the overall fit to the data as seen in \reffig{chi2}. We also verified that the overall size of the error bars does not depend on the number of PCs for the one-loop case, with the errors being comparable to the errors obtained by \textsc{EE2}. For the $\alpha_{0.01}$ scale cut, we observe in \reffig{s8_dm} no detectable bias for the one-loop EFT. As seen in the upper panel of \reffig{scalecuts} and already suggested by the $\chi^2$ analysis, this scale cut is highly conservative, retaining primarily high-redshift data and excluding most non-linear scales. For $\alpha_{0.05}$, the one-loop prediction remains unbiased up to $k_{\rm UV} = 0.5\,h/\mathrm{Mpc}$. However, as $\kmax$ increases, noticeable fluctuations develop within the $0.3\sigma$ band, indicating that the PCA part of the counterterms alone are insufficient to absorb contributions from increasingly non-linear scales as suggested by the sharp $\chi^2$ increase after $k_{\rm UV} = 0.3\,h/\mathrm{Mpc}$. 

The two-loop prediction yields further significant improvement in terms of the scales which can be reached. The right panels of \reffig{s8_dm} represent the two-loop results for different numbers of principal components. The error bars are shown for the case with five PCs, which we verified increases the error bar on average by about $30\%$ compared to the case without any extra components, but, as discussed above, substantially reduces the $\chi^2$. The extra five PCs also make the error bar for the two-loop EFT about $30\%$ larger than for \textsc{EE2} and the one-loop EFT with one PC. 
The need for extra PCs becomes clearer when baryons are added to the datavector, as we discuss in \refsec{baryons}, where the two-loop result seem more robust to baryonic feedback and provides less biased constraints on smaller scales than \textsc{EE2} and one-loop EFT.

For the $\alpha_{0.01}$ scale cut, the bias n the two-loop EFT is essentially flat and consistent with zero across the full range of $\kmax$ considered. In this case as well, adding extra PCs does not lead to any noticeable improvement. For $\alpha_{0.05}$, the bias remains approximately flat up to $k_{\rm UV} = 0.4\,h/\mathrm{Mpc}$ when no additional PCs are included, and stays unbiased up to $k_{\rm UV} \simeq 0.5\,h/\mathrm{Mpc}$. In this case, we observe a significant improvement when additional PCs are introduced which allow for an unbiased recovery of $S_8$ up to $k_{\rm UV} = 0.7\,h/\mathrm{Mpc}$. However, we stress the possibility of overfitting in this regime once more.

\subsection{Sensitivity to baryonic physics} \label{sec:baryons}

A major advantage of the EFTofLSS framework is that it marginalizes over baryonic effects without assuming a specific feedback model or astrophysical inputs, while also avoiding the need to run hydrodynamic simulations \cite{Lewandowski:2014rca,Braganca:2020nhv}. Alternatively, as considered here, such inputs can be used to inform priors on the counterterm parameters. Updated constraints on baryonic feedback~\cite{Xu:2025evn} indicate that, at $k=1\,h/$Mpc, one may expect up to a $10\%$ suppression of the matter power spectrum at $z\leq1$ relative to the dark matter–only case. This effect is particularly relevant for weak lensing, where the projected field receives contributions from all $k$-modes, including those dominated by baryonic physics.

In this section, we contaminate the cosmic shear computed for the fiducial cosmology in \reftab{fiducial_cosmo} using two baryonic models, \textsc{BAHAMAS} and \textsc{OWLS-AGN}, the latter exhibiting stronger feedback than the former. The cosmic-shear $\chi^2$ as a function of $\kmax$ is shown for both feedback models in the middle and right panels of \reffig{chi2}, for $\alpha_{0.01}$ (first row) and $\alpha_{0.05}$ (second row). In these cases, the $\chi^2$ for the different models should be interpreted with two effects in mind: how well a model can accommodate baryonic physics and how well it captures non-linear clustering up to a given scale, with \textsc{EE2} fully addressing the latter but not the former. The comparison with the dark matter–only case then allows one to understand how well each model accommodates baryonic effects.

In the conservative case $\alpha_{0.01}$, \textsc{EE2} accommodates the \textsc{BAHAMAS} feedback well.
For $\alpha_{0.05}$, which allows greater leakage from UV modes where baryonic feedback is more significant, \textsc{EE2} already fails at $k_{\rm UV} = 0.25\,h/\mathrm{Mpc}$. 

When considering $1\%$ UV-contamination case, the one-loop EFT modelling provides a good fit, but slightly worse than \textsc{EE2}, indicating non-linearities in the dark matter part dominate this regime. When allowing for $5\%$ UV contamination, it reaches $\chi^2 \sim 1$ already at $k_{\rm UV} = 0.2\,h/\mathrm{Mpc}$ for both \textsc{BAHAMAS} and \textsc{OWLS-AGN}.
Similar to the dark matter-only case, we find no improvement in the one-loop results when increasing the number of PCs considered. 

As discussed in \refsec{priors}, the priors on the one-loop PCs are somewhat narrow for \textsc{OWLS-AGN}, as illustrated in \reffig{priors} for $k_{\rm UV} = 0.5\,h/\mathrm{Mpc}$ and $\alpha_{0.05}$. However, we observe no significant difference between the one-loop results for \textsc{OWLS-AGN} and \textsc{BAHAMAS}, since both lie within the prior range shown in \reffig{priors}. This indicates that the failure of the one-loop model on these scales is driven by its limited treatment of non-linearities, as evidenced in the left panel of \reffig{chi2}, rather than by overly restrictive priors.

For the two-loop EFT, shown in {orange in} \reffig{chi2}, the results are the most promising, yielding a better fit than all other power-spectrum models. 
This is not suprising when comparing to the one-loop result, which has fewer counterterms and a more restricted pertrubative reach. The fact that the two-loop perforems better than \textsc{EE2} indicates that the two-loop is able to absorb non-linearities as well as baryons, the latter not modelled with \textsc{EE2}.
We also observe a significant improvement for both \textsc{OWLS-AGN} and \textsc{BAHAMAS} when additional PCs are included in the two-loop analysis. This improvement persists over a broad range of $\kmax$, for both $\alpha_{0.01}$ and $\alpha_{0.05}$, and appears at lower $\kmax$ than in the dark matter–only case (left panels).

In \reffig{s8_baryons}, we present the forecasted constraints on $S_8$ for \textsc{OWLS-AGN} (triangle markers with '\textbackslash' hatching), with $\log_{10}[T_{\rm AGN}/(1 \text{K})] = 8.0$ \cite{Le_Brun_2014}, and for \textsc{BAHAMAS} (circle markers with '/' hatching), with $\log_{10}[T_{\rm AGN}/(1 \text{K})] = 7.8$~\cite{McCarthy:2016mry,McCarthy_2018}. We fix the number of PCs to one in the one-loop case and to five in the two-loop case. The dark matter–only results from \reffig{s8_dm} are shown as dashed lines for comparison. The top row displays the constraints for $\alpha_{0.01}$, while the bottom row shows those for $\alpha_{0.05}$.

When \textsc{BAHAMAS} feedback is included, none of the three models exhibits a bias in $S_8$ for $\alpha_{0.01}$, consistent with the low $\chi^2$ values shown in \reffig{chi2}. Although all models recover an unbiased $S_8$ in the presence of \textsc{BAHAMAS} feedback, the two-loop EFT yields the most stable estimate, displaying negligible deviations from the fiducial $S_8$. This indicates that it is more robust to baryonic feedback than either \textsc{EE2} or the one-loop EFT.
When considering \textsc{OWLS-AGN} feedback with $\alpha_{0.01}$, the \textsc{EE2} model, which does not include baryonic effects, breaks down at $\kmax = 0.35\,h/\mathrm{Mpc}$, while the one-loop EFT fails at $\kmax = 0.3\,h/\mathrm{Mpc}$. In contrast, the two-loop EFT remains unbiased up to $\kmax = 0.7\,h/\mathrm{Mpc}$.

When more aggressive cuts are adopted, allowing for $5\%$ UV contamination ($\alpha_{0.05}$), the differences between the non-linear modelling strategies become more pronounced. For \textsc{BAHAMAS} feedback, \textsc{EE2} and the one-loop EFT yield biased estimates of $S_8$ starting at $\kmax = 0.35\,h/\mathrm{Mpc}$ and $\kmax = 0.25\,h/\mathrm{Mpc}$, respectively. In contrast, \reffig{s8_baryons} shows that the two-loop EFT only becomes biased around $\kmax = 0.6\,h/\mathrm{Mpc}$. As discussed in \refsec{darkmatter}, results beyond $\kmax = 0.5\,h/\mathrm{Mpc}$ for the two-loop EFT may already be affected by some degree of overfitting.
When the stronger \textsc{OWLS-AGN} feedback model is considered, the two-loop EFT is the only power-spectrum model that successfully recovers the fiducial $S_8$ value from intermediate $k$ values. In this case, the scale at which the inference reaches a $0.3\sigma$ bias reduces to $\kmax = 0.35\,h/\mathrm{Mpc}$, although this does not correspond to a sharp breakdown.

We therefore find that the two-loop reaches smaller scales than the one-loop EFT, while remaining robust with respect to baryonic feedback. As shown in \reffig{s8_baryons}, the expected error bars for the two-loop EFT 
at a fixed $\kmax$ are generally about $20\%$ larger than in the one-loop case for $\alpha_{0.01}$ and very comparable for $\alpha_{0.05}$, despite the larger number of counterterms (or, in our analysis, the increased number of PCs). The possibility of extending to smaller scales more than compensates for the small loss in terms of constraining power, eventually providing similar or tighter constraints than the other power spectrum models.
Furthermore, while the dark matter-only model with \textsc{EE2} reaches slightly smaller scales than the one-loop EFT, it struggles to reach the same scales as the two-loop EFT.

\section{Conclusion} \label{sec:conclusion}

This paper presents the first application of the two-loop EFTofLSS to galaxy weak lensing. We outline a framework to model the redshift evolution of the counterterms, pre-calibrating them in dark matter-only simulations while retaining sufficient freedom to account for corrections induced by small-scale physics such as  baryons. By performing a PCA on the counterterms, we restrict the parameters to the subspace that the data is sensitive to. 
Recent works pointed out the existence of strong baryonic feedback \cite{Xu:2025evn,Amon:2022azi,Siegel:2025ivd,Hadzhiyska:2024qsl,McCarthy:2024tvp,DES:2024iny,Bigwood:2025kur}, indicating the importance to have a non-linear modelling that is able to capture different feedback features in the observables. Our approach preserves the constraining power of the two-loop EFT while enough flexibility to absorb a range of baryonic feedback effects via the PCA amplitudes.

We derive conservative priors in the PCA-defined parameter space using a variety of hydrodynamical simulations. We highlight our simulation-based prior as a preliminary step towards understanding the connection between different baryonic parametrizations \cite{Schneider:2015wta,Mead:2020vgs,Mead_2021,Aric__2021,Arico:2023ocu} or hydrodynamical simulations \cite{McCarthy_2018,Salcido_2023,Le_Brun_2014, Genel_2014,Vogelsberger_2014,Pillepich_2017,Springel_2017,Naiman_2018,Marinacci_2018,Nelson_2017,Crain_2015,Schaye_2014,Khandai_2015,Chisari_2018} and the EFT counterterms in the context of weak lensing. A better understanding of the mapping between baryonic effects and counterterms may provide insight into baryonic physics, for example by putting constraints on $T_{\rm AGN}$, or lead to more informative priors in the counterterms. We plan to investigate this connection further in future work.

Furthermore, we apply the UV-IR decomposition method inspired by~\cite{derose2025lensingcounternarrativeeffective} to translate scale cuts in $\kmax$ into corresponding cuts in angular scales for cosmic shear. We adopt \textsc{EE2} as a UV completion for the matter power spectrum and present our results in terms of the maximal allowed UV contamination, parameterized by $\alpha$ in \refeq{criteriacut}, verifying that this procedure yields unbiased constraints when validated against simulations. It would be interesting to further explore alternative analytic approaches to UV completion, such as Vlasov perturbation theory~\cite{Garny:2022kbk,Garny:2022tlk}, removing the need for simulation inputs in the UV.

Our results demonstrate that the two-loop EFT analysis for Roman yields unbiased and baryon-modelling-independent constraints on $S_8$ at the level of $0.9\%$ and $1.4\%$, assuming $\kmax = 0.5\,h/\mathrm{Mpc}$, for $\alpha_{0.05}$ and $\alpha_{0.01}$, respectively. 
The one-loop EFT also achieves a $1\%$ constraint on $S_8$, albeit over a more limited perturbative range of $\kmax = 0.2\,h/\mathrm{Mpc}$ for $\alpha_{0.05}$. Allowing for $1\%$ UV contamination ($\alpha_{0.01}$) extends the reach to $\kmax = 0.35\,h/\mathrm{Mpc}$, yielding a $1.3\%$ constraint. The extended range of validity of the two-loop EFT may also be very useful to test dark matter properties that could only set in on smaller scales using galaxy lensing, such as interactions with or decays into other species.

One caveat of our analysis is the use of the \textsc{NLA} model for intrinsic alignments, which has been shown to break down at smaller $k$ than the EFT of galaxy shapes~\cite{Bakx:2023mld,Maion:2023vdf,Chen:2023yyb}. 
This limitation could impact the scalecuts needed for an analysis with real data.
Integrating our two-loop modelling for the non-linear clustering with the EFT of galaxy shapes~\cite{Vlah:2019byq,Chen:2023yyb} provides an interesting next step that would clarify the usable scales for galaxy lensing.

Combining multiple probes, such as cosmic shear, galaxy clustering, and CMB lensing, within a unified EFT-based framework is a particularly promising short-term direction. In this work, we provide the first step toward a two-loop implementation for the shear-shear correlation. While extending the two-loop framework to biased tracers introduces a substantial number of additional free parameters \cite{Schmidt:2020tao, Bakx:2025cvu, Donath:2023sav}, the application to the matter field requires only five redshift-dependent counterterms, shared across different observables. As a result, combining multiple datasets spanning a range of redshifts may yield stringent and internally consistent constraints within a single theoretical framework, further breaking degeneracies between counterterms and cosmological parameters.

\section*{Acknowledgements}

The authors would like to thank Stony Brook Research Computing and Cyberinfrastructure and the Institute for Advanced Computational Science at Stony Brook University for access to the high-performance SeaWulf computing system, which was made possible by $1.85$M in grants from the National Science Foundation (awards 1531492 and 2215987) and matching funds from the Empire State Development’s Division of Science, Technology and Innovation (NYSTAR) program (contract C210148). VM \& TE are supported by the Roman Project Infrastructure Team ``Maximizing Cosmological Science with the Roman High Latitude Imaging Survey" (NASA contracts 80NM0018D0004-80NM0024F0012). VM is also partially supported by the Roman Project Infrastructure Team ``A Roman Project Infrastructure Team to Support Cosmological Measurements with Type Ia Supernovae" (NASA contract 80NSSC24M0023). 

\appendix

\section{Loop Integral Emulator} \label{app:Loop_emulator}

\begin{figure}[!t]
    \centering
    \includegraphics[width=\columnwidth]{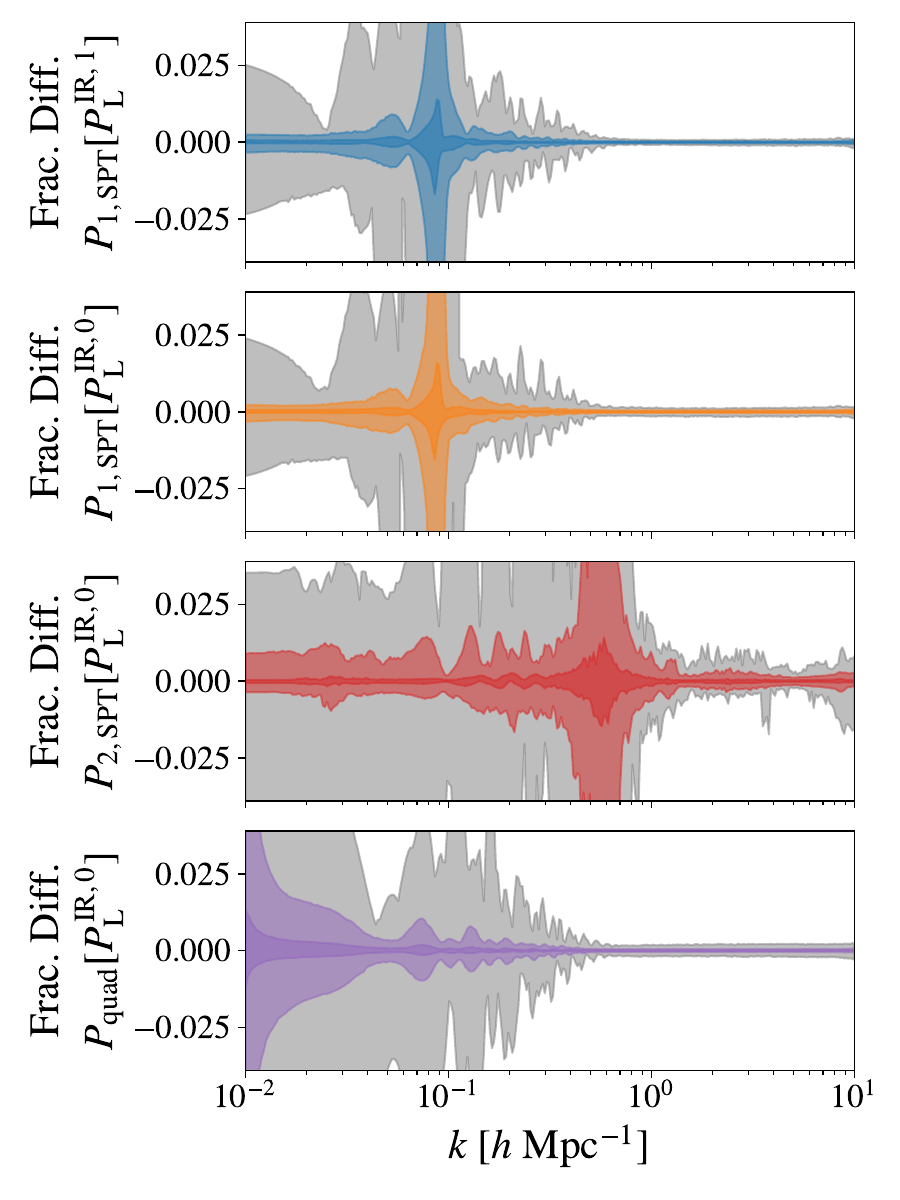}
    \caption{The fractional difference of each of the loop integrals needed for our analysis. The colored regions show the $67$\% and $95$\% percentiles, while the gray region shows the maximum observed emulation error. The superscript corresponds to the loop order that the term is used for in accordance with the IR-resummation scheme we implement, which is outlined in~\cite{Bakx:2025jwa}. The sharp rise in the error near $k=0.1 \,h/$Mpc for the one-loop, $k=0.6\,h/$Mpc for the two-loop, and $k=0.01\, h/$Mpc for the quadratic counterterm, are caused by the loop integral being close to zero rather than emulation error. With this, the loop emulators are almost always at the sub-percent level error, the same order as the requested numerical error of $0.1\%$.}
    \label{fig:loop_emulator_errors}
\end{figure}

The two-loop corrections require integration of the two internal momenta, giving rise to a five dimensional integral. Reaching sub-percent accuracy of the integrals takes $\mathcal{O}(1-10)$ hours when evaluating using the \textsc{SUAVE} algorithm in \textsc{Cuba}~\cite{Hahn_2005} when utilizing $\mathcal{O}(10-100)$ CPU cores, depending on the specific precision and number of cores. We opt then to emulate the boost for the individual loop integrals as a function of the cosmological parameters. Specifically, we train a neural network (NN) to learn the mapping
\begin{equation}\label{eq:loop_emulator}
    (A_s \times 10^9, n_s, H_0, \Omega_b, \Omega_m ) \rightarrow \sinh^{-1}\left(\frac{X(k)}{P_{\rm L}(k)}\right)\,,
\end{equation}
where 
\begin{equation}
\begin{split}
    X\in\bigg\{&
          P_{\rm 1,SPT} \left[ P_{\rm L}^{{\rm IR},1} \right],
        \,P_{\rm 1,SPT} \left[ P_{\rm L}^{{\rm IR},0} \right],\\
    &   \quad\quad
          P_{\rm 2,SPT} \left[ P_{\rm L}^{{\rm IR},0} \right],
        \,P_{\rm quad}  \left[ P_{\rm L}^{{\rm IR},0} \right] \bigg\} \,,
\end{split}
\end{equation}
is the loop integral to emulate. The $P_{\rm L}^{{\rm IR},i}$ corresponds to the IR resummation following the notation in~\cite{Garny_2022}. We evaluated each of the $X$ at $300$ $k$-modes log-evenly distributed in the range $[10^{-2}, 10^{1}]$ and at $1000$ different cosmologies randomly sampled from the CMB first acoustic peak posterior (see \refsec{priors} for details) to be used as the training data. This allows one to treat the output functions in the righthand side of \refeq{loop_emulator} as $300$ dimensional vectors.

\begin{figure*}[t!]
    \centering
    \includegraphics[width=\textwidth]{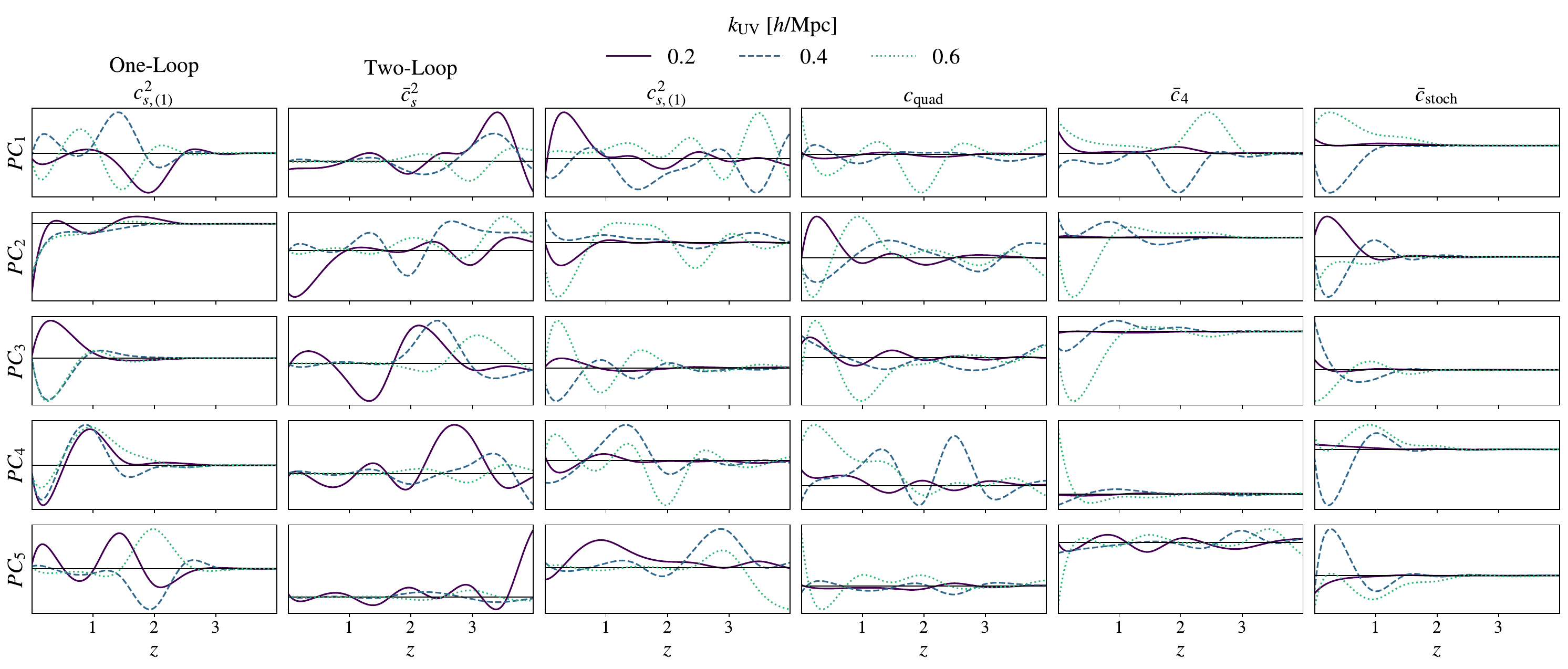}
    \caption{
    The first five principal components, each normalized to unity, for the EFT counterterms for different values of $\kmax$ when using the $\alpha_{0.05}$ cuts. The leftmost column is for the one-loop counterterm $c_{s,(1)}^2(z)$, while the rest of the columns are for the two-loop counterterms $\vec{c}(z)$. The solid purple lines correspond to $\kmax=0.2\,h/{\rm Mpc}$, the dashed blue to  $\kmax=0.4\,h/{\rm Mpc}$, and the dotted green to $\kmax=0.6\,h/{\rm Mpc}$. 
    }
    \label{fig:pca}
\end{figure*}

To reduce the dimensionality and improve NN performance, we perform a PCA on the output vectors to extract an orthonormal basis that represent the features of maximal variance across cosmological parameters. We train an NN to predict the first $8$, $8$, $10$, and $6$ PCA amplitudes for each of the $X$, respectively, with each $X$ getting its own NN.
We use the \textsc{ResMLP} architecture outlined in~\cite{zhong2024attentionbasedneuralnetworkemulators,saraivanov2024attentionbasedneuralnetworkemulators,zhu2025attentionbasedneuralnetworkemulators} with three \textsc{ResBlocks} with layer widths of $128$, $128$, $512$, and $128$ for each of the $X$, respectively. The NNs are optimized using the mean square error of the PCA amplitudes as the loss function. Every layer is followed by the activation function defined in~\cite{Alsing_2020,Spurio_Mancini_2022,zhu2025attentionbasedneuralnetworkemulators}. We do not use any regularization or dropout since we do not observe overfitting of the training data.

The fractional error between the emulated and numerical loop integrals are presented in \reffig{loop_emulator_errors}. Our procedure accurately emulates $95$\% of cosmologies within the CMB first acoustic peak prior (see \refsec{priors} for details) at the sub-percent level, and $100$\% of cosmologies at the percent level. The spikes in the fractional error of the emulators occur when $X$ is small, meaning minor absolute errors become large fractional errors. However, these have negligible impact on the total $P_{\rm NL}(k)$.

\section{Details of the PCA} \label{app:PCA}

In \reffig{pca}, we show the first five principal components for the one and two-loop PCA decomposition of the counterterms, each normalized to unity, for a few values of $\kmax$ using the $\alpha_{0.05}$ cuts. The details of the PCA are discussed in \refsec{priors}. The one-loop PCA basis is shown in the leftmost column, and the remaining columns show the two-loop PCA basis.

\bibliographystyle{apsrev4-2}
\bibliography{bib}

\end{document}